\documentclass[times, twocolumn]{aastex631}
\usepackage{graphicx, graphics}
\usepackage{CJK}
\usepackage{bbm}
\usepackage{comment}
\definecolor{cyan(process)}{rgb}{0.0, 0.72, 0.92}

\usepackage[T1]{fontenc} % enable polish characters for Lau 2022 citation

\usepackage{amsmath, amssymb, bm}
\usepackage[nodayofweek]{datetime}
\usepackage{multirow}
\input{prepcitationformat.txt}

\makeatletter
\newcommand*{\centerfloat}{%
  \parindent \z@
  \leftskip \z@ \@plus 1fil \@minus \textwidth
  \rightskip\leftskip
  \parfillskip \z@skip}
\makeatother

\newdateformat{monthyearday}{%
  \THEYEAR\ \monthname[\THEMONTH] \twodigit{\THEDAY}}

\newcommand{\uat}[2]{\href{http://vocabs.ands.org.au/repository/api/lda/aas/the-unified-astronomy-thesaurus/current/resource.html?uri=http://astrothesaurus.org/uat/#1}{#2 (#1)}}

\graphicspath{{./}{figures/}}

\newcommand{\kms}{\hbox{km s$^{-1}$}}

\newcommand{\vsini}{\hbox{$v \sin i$}}

\shorttitle{Twenty-Five Years of Accretion onto TW Hya}
\shortauthors{Herczeg, Chen, et al.}

\usepackage{CJK}
\begin{document}
\begin{CJK*}{UTF8}{gbsn}

% Title.
\title{Twenty-Five Years of Accretion onto the Classical T Tauri Star TW Hya}

\author[0000-0002-7154-6065]{Gregory J. Herczeg (沈雷歌)}
\affiliation{Kavli Institute for Astronomy and Astrophysics, Peking University, Beijing 100871, China}
\affiliation{Department of Astronomy, Peking University, Beijing 100871, China}
\affil{Visiting astronomer, Department of Astronomy; California Institute of Technology; Pasadena, CA 91125, USA}

\author[0000-0003-4520-5395]{Yuguang Chen (陈昱光)}
\affiliation{Department of Physics \& Astronomy, University of California, Davis, 1 Sheilds Avenue, Davis, 95616, CA, USA}

\author[0000-0001-5541-2887]{Jean-Francois Donati}
\affiliation{IRAP-UMR 5277, CNRS \& Univ. de Toulouse, 14 Av. E. Belin, F-31400, Toulouse, France}

\author[0000-0002-8985-8489]{Andrea K. Dupree}
\affiliation{Center for Astrophysics | Harvard \& Smithsonian}

\author[0000-0001-7796-1756]{Frederick M. Walter}
\affil{Department of Physics \& Astronomy, Stony Brook University,
             Stony Brook NY 11794-3800, USA}

\author{Lynne A. Hillenbrand}
\affiliation{Department of Astronomy; California Institute of Technology; Pasadena, CA 91125, USA}

\author[0000-0002-8828-6386]{Christopher M. Johns-Krull}
\affil{Department of Physics and Astronomy, Rice University, 6100 Main Street, Houston, TX 77005, USA}

\author[0000-0003-3562-262X]{Carlo F. Manara}
\affil{European Southern Observatory, Karl-Schwarzschild-Strasse 2, 85748 Garching bei M\"unchen, Germany}

\author[0000-0003-4243-2840]{Hans Moritz G{\"u}nther}
\affiliation{MIT Kavli Institute for Astrophysics and Space Research, 77 Massachusetts Avenue, Cambridge, MA 02139, USA}

\author[0000-0001-8060-1321]{Min Fang (房敏)}
\affiliation{Purple Mountain Observatory, Chinese Academy of Sciences, 10 Yuanhua Road, Nanjing 210023, China}
\affiliation{University of Science and Technology of China, Hefei 230026, China}

\author[0000-0002-5094-2245]{P. Christian Schneider}
\affiliation{Hamburger Sternwarte, Gojenbergsweg 112, 21029, Hamburg, Germany}

\author[0000-0003-3305-6281]{Jeff A. Valenti}
\affil{Space Telescope Science Institute, 3700 San Martin Drive, Baltimore, MD 21218, USA}

\author{Silvia H.P. Alencar}
\affil{Departamento de Fisica - ICEx - UFMG, Av. Antonio Carlos 6627, 30270-901 Belo Horizonte, MG, Brazil}

\author{Laura Venuti}
\affil{SETI Institute, 339 Bernardo Ave., Suite 200, Mountain View, CA 94043, USA}

\author[0000-0001-8657-095X]{Juan Manuel Alcal\'a}
\affil{INAF-Osservatorio Astronomico di Capodimonte, via Moiariello 16, 80131 Napoli, Italy}

\author[0000-0002-0474-0896]{Antonio Frasca}
\affil{INAF - Osservatorio Astrofisico di Catania, via S. Sofia, 78, 95123 Catania, Italy}

\author[0000-0003-2631-5265]{Nicole Arulanantham}
\affil{Space Telescope Science Institute, 3700 San Martin Drive, Baltimore, MD 21218, USA}

\author[0000-0003-4446-3181]{Jeffrey L. Linsky}
\affil{JILA, University of Colorado, Boulder CO 80309-0440}

\author{Jerome Bouvier}
\affil{Univ. Grenoble Alpes, CNRS, IPAG, 38000 Grenoble, France}

\author[0000-0002-8704-4473]{Nancy S. Brickhouse}
\affiliation{Center for Astrophysics | Harvard \& Smithsonian}

\author{Nuria Calvet}
\affiliation{Department of Astronomy, University of Michigan, 1085 South University Avenue, Ann Arbor, MI 48109, USA}

\author[0000-0001-9227-5949]{Catherine C. Espaillat}
\affiliation{Institute for Astrophysical Research, Department of Astronomy, Boston University, 725 Commonwealth Avenue, Boston, MA 02215, USA}

\author[0000-0002-3913-3746]{Justyn Campbell-White}
\affil{European Southern Observatory, Karl-Schwarzschild-Strasse 2, 85748 Garching bei M\"unchen, Germany}

\author[0000-0003-2251-0602]{John M. Carpenter}
\affiliation{Joint ALMA Observatory, Avenida Alonso de C\'ordova 3107, Vitacura, Santiago, Chile}

\author[0000-0002-0112-5900]{Seok-Jun Chang}
\affiliation{Max-Planck-Institut f\"{u}r Astrophysik, Karl-Schwarzschild-Strasse 1, 85748 Garching b. M\"{u}nchen}

\author[0000-0002-1821-0650]{Kelle L. Cruz}
\affiliation{Dept of Physics and Astronomy, Hunter College, City University of New York, NY, NY, USA}
\affiliation{Physics, Graduate Center of the City University of New York, NY, NY, USA}
\affiliation{Department of Astrophysics, American Museum of Natural History, Ny, NY, USA}

\author{S.E. Dahm}
\affiliation{Gemini Observatory/NSF’s NOIRLab, 950 N. Cherry Avenue, Tucson, AZ, 85719, USA}

\author[0000-0001-6496-0252]{Jochen Eisl\"offel}
\affiliation{Th\"uringer Landessternwarte, Sternwarte 5, D-07778
Tautenburg, Germany}

\author[0000-0002-3232-665X]{Suzan Edwards}
\affiliation{Five College Astronomy Department, Smith College, Northampton, MA 01063, USA}

\author[0000-0002-3747-2496]{William J. Fischer}
\affil{Space Telescope Science Institute, 3700 San Martin Drive, Baltimore, MD 21218, USA}

\author[0000-0003-0292-4832]{Zhen Guo (郭震)}
\affil{Instituto de F{\'i}sica y Astronom{\'i}a, Universidad de Valpara{\'i}so, ave. Gran Breta{\~n}a, 1111, Casilla 5030, Valpara{\'i}so, Chile}
\affil{N\'ucleo Milenio de Formaci\'on Planetaria (NPF), Chile, ave. Gran Breta{\~n}a, 1111, Casilla 5030, Valpara{\'i}so, Chile}
\affil{Centre for Astrophysics Research, University of Hertfordshire, Hatfield AL10 9AB, UK}
\affil{Departamento de F{\'i}sica, Universidad Tecnic{\'a} Federico Santa Mar{\'i}a, Avenida Espa{\~n}a 1680, Valpara{\'i}so, Chile}

\author{Thomas Henning}
\affil{Max-Planck-Institut f\"ur Astronomie, K\"onigstuhl 17, D-69117 Heidelberg, Germany}

\author{Tao Ji (纪涛)}
\affiliation{Kavli Institute for Astronomy and Astrophysics, Peking University, Beijing 100871, China}
\affiliation{Department of Astronomy, Peking University, Beijing 100871, China}

\author[0000-0003-4908-4404]{Jessy Jose}
\affiliation{Indian Institute of Science Education and Research (IISER) Tirupati, Rami Reddy Nagar, Karakambadi Road, Mangalam (P.O.), Tirupati 517507, India}

\author[0000-0002-3138-8250]{Joel H. Kastner}
\affil{Center for Imaging Science, School of Physics and Astronomy, and Laboratory for Multiwavelength Astrophysics, Rochester Institute of Technology, Rochester, NY 14623, USA}

\author[0000-0002-8298-2663]{Ralf Launhardt}
\affil{Max-Planck-Institut f\"ur Astronomie, K\"onigstuhl 17, D-69117 Heidelberg, Germany}

\author[0000-0002-7939-377X]{David A. Principe}
\affil{MIT Kavli Institute for Astrophysics and Space Research, 77 Massachusetts Avenue, Cambridge, MA 02139, USA}

\author[0000-0003-1639-510X]{Connor E. Robinson}
\affil{Department of Physics \& Astronomy, Amherst College, C025 Science Center 25 East Drive, Amherst, MA 01002, USA}

\author[0000-0001-7351-6540]{Javier Serna}
\affil{Instituto de Astronom\'{i}a, Universidad Nacional Aut\'{o}noma de M\'{e}xico Ensenada, B.C, M\'{e}xico}

\author[0000-0002-0786-7307]{Micha{\l} Siwak}
\affiliation{Konkoly Observatory, Research Centre for Astronomy and Earth Sciences, E{\"o}tv{\"o}s Lor\'and Research Network (ELKH), Hungarian Academy of Sciences, Konkoly-Thege Mikl\'os \'ut 15--17, 1121 Budapest, Hungary}
\affiliation{CSFK, MTA Centre of Excellence, Budapest, Konkoly Thege Mikl\'os \'ut 15-17, H-1121, Hungary}

\author[0000-0002-5784-4437]{Michael F. Sterzik}
\affil{European Southern Observatory, Karl-Schwarzschild-Strasse 2, 85748 Garching bei M\"unchen, Germany}

\author[0000-0003-3882-3945]{Shinsuke Takasao}
\affiliation{Department of Earth and Space Science, Graduate School of Science, Osaka University, Toyonaka, Osaka 560-0043, Japan}

\correspondingauthor{Gregory Herczeg, Yuguang Chen} 
\email{gherczeg1@gmail.com, yugchen@ucdavis.edu}

\begin{abstract}
Accretion plays a central role in the physics that governs the evolution and dispersal of protoplanetary disks.  The primary goal of this paper is to analyze the stability over time of the mass accretion rate onto TW Hya, the nearest accreting solar-mass young star.  We measure veiling across the optical spectrum in 1169 archival high-resolution spectra of TW Hya, obtained from 1998--2022.  The veiling is then converted to accretion rate using 26 flux-calibrated spectra that cover the Balmer jump.  The accretion rate measured from the excess continuum has an average of $2.51\times10^{-9}$~M$_\odot$~yr$^{-1}$ and a Gaussian distribution with a FWHM of 0.22 dex.  This accretion rate may be underestimated by a factor of up to 1.5 because of uncertainty in the bolometric correction and another factor of 1.7 because of excluding the fraction of accretion energy that escapes in lines, especially Ly$\alpha$.
The accretion luminosities are well correlated with He line luminosities but poorly correlated with H$\alpha$ and H$\beta$ luminosity.  
The accretion rate is always flickering over hours but on longer timescales has been stable over 25 years.
This level of variability is consistent with previous measurements for most, but not all, accreting young stars.
\end{abstract}


\section{Introduction}

The evolution of protoplanetary disks and the final outcome of any planet formation depends in part on how gas flows through the disk \citep[see review by][]{manara22}.  Since the flows within the disk are challenging to measure, we often infer the global flow rate by measuring accretion from the disk onto the star.  The disk-to-star accretion rate appears to vary on all timescales and with a wide range of amplitude \citep[see reviews by][]{hartmann16,fischer22}.

The accretion rate and flow properties can be measured using many emission lines and in continuum emission, both of which are produced as material flows from the disk along the stellar magnetosphere and crashes at the stellar surface.
The energy of the infalling gas is initially deposited at the base of the accretion flow, which is shock-heated to $\sim 10^6$ K \citep[e.g.][]{Calvet1998,lamzin98}.  The accretion shock itself occurs near the 
stellar surface and heats the surrounding photosphere to $\sim 10^4$ K  \citep[e.g.][]{Drake2005,Brickhouse2012,Bonito2014}.  Most of the accretion energy is reprocessed and escapes as hydrogen continuum emission from the heated photosphere \citep[e.g.][]{Calvet1998}, which may then be used to measure accretion rates after scaling by a bolometric correction \citep[e.g.][]{Valenti1993,Hartigan1995,Gullbring1998}.  The funnel flows and accretion shock also produce line emission \citep[e.g.][]{Muzerolle2000,Kurosawa2006,Donati2014}, which may be converted into an accretion luminosity or accretion rate using correlations with the accretion continuum measurements \citep[e.g.][]{Natta2004,Fang2009,alcala17}.  

Each of the observable diagnostics of accretion rate has its own advantages and disadvantages.    Photometry covers large samples on short and long timescales, though changes in accretion are difficult to unambiguously distinguish from other phenomena, such as chromospheric flares, changing spot coverage fractions, and extinction  \citep[e.g.,][]{bouvier93,cody14,hillenbrand22}.
Flux-calibrated spectra covering the Balmer jump provide an instantaneous measurement of accretion, but with repeated observations for only a few objects \citep[e.g.][]{robinson19}. Multi-epoch measurements of veiling in optical high-resolution spectra provide a consistent set of accretion rates, assuming that the underlying photospheric emission remains constant
 \citep[e.g.][]{johnskrull97,alencar12}.  
Variability may also be inferred from changes in line emission \citep[e.g.][]{scholz05,Costigan2014}, although for one accreting young star, XX Cha, large changes in the accretion continuum did not result in significant changes in line luminosities \citep{claes22}.
Some recent work has combined spectroscopy with extensive photometric monitoring to interpret the photometric changes with more precision and across longer periods of time \citep[e.g.][]{venuti21,zsidi22,fiorellino22,bouvier20,bouvier23}, with typical changes in accretion rate  of $\sim 0.3$ dex.

\begin{table*}[!t]
\centering
\caption{Summary of Instruments}
\begin{tabular}{llccccccr}
\hline\hline
Telescope & Instrument & $\lambda$ range (\AA) & Resolution & Aperture & Sky Subtr. & \#  & Years & Instr. Reference\\
\hline
\multicolumn{9}{c}{High Resolution Spectra}\\
\hline
\hline
CFHT & ESPaDOnS    & 3800--10000 & 68,000 & $1\farcs7$ fiber & No & 284  & 2008--2016 & \citet{donati06} \\
\hline
Magellan & MIKE-red$^a$ & 4850--9100 & 35,000 & $0\farcs75$ slit & Yes & 467 & 2004--2007 & \citet{bernstein03} \\
ESO-MPG 2.2m & FEROS        &  3800--9200 & 48,000 & $2^{\prime\prime}$ fiber & Yes & 241      & 1998--2019 & \citet{kaufer99} \\
ESO 3.6m & HARPS   &  3800--6900  &115,000 & $1^{\prime\prime}$ fiber & No & 34 &2005--2009& \citet{mayor03}\\
VLT & UVES$^a$         &  5000--7000$^a$ & 57,000$^a$ & $0\farcs7^a$ slit & Yes & 47 & 2000--2022 & \citet{dekker00}\\
VLT & ESPRESSO     &  3800--7900 & 140,000 & $1^{\prime\prime}$ fiber & Yes & 5 & 2021 & \citet{pepe10}\\
SMARTS & CHIRON    &  4100--8900 & 30,000 & $2\farcs7$ fiber &  No & 68 & 2021--2022  & \citet{tokovinin13}\\
Keck I & HIRES$^a$ & 4800--9200$^a$ & 36,000$^a$ & $1.15^a$ slit & Yes & 18 & 2000--2015 &  \citet{vogt94} \\
McDonald 2.7m & 2d-coude & 4000-10000 & 50,000 & $1.15$ slit & Yes & 4 & 2006 & \citet{tull95} \\
CTIO 4m & Echelle  & 4375-7600 & 25,000 & $1.5^{\prime\prime}$ slit & Yes & 1 & 1998 &  --\\
\hline
\multicolumn{9}{c}{Low and medium Resolution Spectra}\\
\hline
\hline
VLT & X-Shooter & 3000-25000 & $10,000^a$ & $1\farcs0$ slit$^a$ & Yes & 6 & 2010-2021 & \citet{vernet11}\\
UH88 & SNIFS & 3100--10000 & $1000$ & IFU & Yes & 1 & 2014 & \citet{lantz04}\\
HST & STIS & 3000--5700 & 1000 & $0\farcs2$ slit & Yes & 4 & 2000-2015 & \citet{woodgate98}\\
Keck & LRIS & 3100--9500 & 1000 & $1\farcs0$ slit & Yes & 1 & 2008 & \citet{oke95}\\
Palomar & DBSP & 3200--9000 & 1000 & $2\farcs0$ slit & Yes & 6 & 2008 & \citet{oke82}\\
\hline
\multicolumn{8}{l}{$^a$Typical setting, some spectra obtained with other wavelength settings and slit widths}\\
\end{tabular}
\label{observation}
\end{table*}

In this paper, we contribute to these efforts to understand accretion variability by analzying 1169 high-resolution spectra of the classical T Tauri star (CTTS) TW Hya, obtained over 25 years.  
As one of the closest and brightest CTTSs, TW Hya has been repeatedly observed because it is a remarkable object, a cornerstone for studies of accretion, and a common target of radial velocity searches for young exoplanets.
We compile and analyze 1169 high-resolution optical spectra from CFHT/ESPaDOnS,  MPG-ESO 2.2m/FEROS, Magellan/MIKE, La Silla 3.6m/HARPS, VLT/UVES, VLT/ESPRESSO, Keck/HIRES, McDonald 2.7m/2coude, and CTIO 4m/cassegrain echelle to conduct the largest analysis of veiling for a single star.  The veiling measurements are then converted to accretion luminosity, guided by $26$ flux-calibrated spectra, most obtained at low-resolution.  The paper is organized as follows: \S2 describes the properties of TW Hya; \S3 introduces the archival data and data reduction process; \S4 describes the veiling decomposition technique and how veiling is converted into accretion rate; \S5 describes our use of low-resolution spectra to convert veiling to accretion rate; \S6 compares veiling measurements to emission line diagnostics of accretion; and finally \S7 discusses the accretion variability and searches for timescales.

\section{Properties of TW Hya}

TW Hya was initially discovered as an emission line object by \citet{henize76} and has since become the namesake of the young, nearby TW Hya Association \citep{delareza89,kastner97,hoff98}.  The disk of TW Hya is one of the few disks in the TW Hya Association \citep[e.g.][]{weinberger13,luhman23}, as it has survived for longer than the typical disk dissipation timescale \citep[e.g.][]{Hernandez2008,mamajek09,Fedele2010}.
The disk is massive and large, with a series of rings and gaps and an inner hole in micron-sized dust \citep[e.g.][]{calvet02,bergin13,andrews16,vanboekel17}.

The spectral type of TW Hya in the literature ranges from K6/K7 when measured in blue/optical wavelengths and M2 when measured in the near-IR  \citep[e.g.][]{webb99,yang05,vacca11,McClure2013,debes13}.  This wavelength dependence is likely explained by spots that cover some of the visible surface \citep[e.g.][]{debes13,gully17,gangi22}.  Since the TW Hya disk and magnetosphere are both viewed nearly pole-on \citep[e.g.][]{qi08,Donati2011}, the visible spot coverage is expected to be roughly steady, without significant rotational modulation (although some photometric and spectroscopic variability is detected and attributed to the spot, see discussion below).  We adopt an intermediate spectral type of M0.5 and a corresponding temperature of $3810$ K, based on analyses of low-resolution optical spectra (\citealt{Herczeg2014}, see also, e.g., \citealt{sokal18} and \citealt{venuti19}).  

The parameters $T_{\rm eff}=3810$ K, $J=8.217\pm0.024$ mag from 2MASS \citep{Cutri2003}, $J$-band bolometric correction from \citet{pecaut13}, an estimated $J$-band veiling of 0.1 \citep{fischer11}, and 
$d=60.14\pm0.05$ pc \citep{gaia21}
yield $\log L_{\rm phot}/L_{\rm \odot}=-0.54$ and radius 1.23~R$_\odot$.  This temperature and luminosity correspond to a mass of 0.87~M$_\odot$ and an age of 9.8 Myr for the \citet{somers20} evolutionary tracks for stars with 50\% spot coverage and 0.59~M$_\odot$ and an age of 4.3 Myr for 0\% spot tracks (the 0\% tracks are similar to the results from the \citealt{Baraffe2015} models).  We adopt the 0.87~M$_\odot$ mass, based on consistency for a few low-mass stars between mass estimates from the 50\% spot models and dynamical masses from the rotation of their disks  \citep{pegues21}.  For the 50\% spot models and corresponding bolometric corrections, a temperature of $\sim 4100$ K would lead to a mass of 0.96 $M_\odot$ and an age of 16 Myr; a lower temperature of $\sim 3600$ K leads to 0.70 M$_\odot$ and 8.5 Myr.

The dynamical mass of $0.81\pm0.17$~M$_\odot$ measured by \citet{teague19} from ALMA gas observations is more consistent with the 50\% spot model than the 0\% spot model.   The adopted mass and radius lead to a surface gravity $\log g$ of  4.2, consistent with a near-IR measurement of surface gravity by \citet{sokal18}, slightly larger than the 4.02 measured by \citet{mentuch08} and lower than 4.46 of \citet{venuti19}, both from optical spectra.
Line emission and absorption in \ion{He}{1} and \ion{He}{2} lines is detected at $\sim 400$--$450$ \kms\ on the red side of the line (see \S 6), which requires a mass larger than $0.8$~M$_\odot$ for free-fall from 5 $R_*$ (or $\sim 0.9$ M$_\odot$ for freefall from 3.5 R$_*$).

X-ray observations reveal the dense shock where the accretion flow heats the star \citep[e.g.][]{Kastner2002,stelzer04,Brickhouse2010,argiroffi17}.  Infall velocities of $\sim 450$ \kms\ are seen in \ion{He}{1} $\lambda10830$ and hot ultraviolet lines  \citep[e.g.][]{Herczeg2002,dupree05,johnskrullherczeg07,Ardila2013}.
The X-ray emission is redshifted by only $38$ \kms, consistent with emission in the post-shock region \citep{argiroffi17}.  
Zeeman Doppler Imaging and polarimetry indicate that the accretion morphology is dipole-like, with polar spots where the flow strikes the stellar surface \citep[e.g.][]{Donati2011,JohnsKrull2013}.  
Previously measured accretion rates from continuum and line analyses range from $4\times10^{-10}-3\times10^{-9}$ M$_\odot$ yr$^{-1}$  \citep[e.g.][]{Muzerolle2000,Alencar2002,Herczeg2002,Donati2011,robinson19}, with differences caused by real variability and also by methodological differences. 

Monitoring in line and continuum emission reveal changes seen on hours and days timescales, usually attributed to variability in accretion \citep[e.g.][]{Alencar2002,Batalha2002,Huelamo2008,Dupree2012}.  The stellar rotation period of 3.568 days is measured from sinusoidal changes in the radial velocity and could indicate the presence of a hot Jupiter \citep{Setiawan2008} but is more likely caused by spot modulation \citep{Huelamo2008}.
High time resolution observations with MOST obtained across several years show frequent bursts on short timescales and no signatures of extinction variability  \citep{Siwak2011,siwak18}, as can be seen for some accreting systems with disks that are viewed at higher inclinations. Flares are also detected in \ion{Si}{4} and \ion{C}{4} lines \citep{hinton22} and in X-rays \citep[e.g.][]{Brickhouse2012}.

Br$\gamma$ emission, produced in the accretion flow, extends across at least 3--4 R$_*$, indicating an inner disk truncation radius of at least that size  \citep{gravity_garcialopez20}, consistent with the truncation radius inferred from accretion rates and the magnetic field strengths \citep{johnskrull07,Donati2011}.  
The disk truncation radius is inside the corotation radius of  $\sim 8$ R$_*$, which corresponds to the 3.56 day period.  In the magnetospheric framework of \citet{Dangelo2010}, the accretion should be stable over the long term, while in the simulations of \citet{Blinova2015}, the accretion onto TW Hya (and most accreting young stars) should be in the ordered unstable regime, with multiple irregular accretion tongues.

\section{Observations and Data Reduction}
\label{sec_observation}

In this section, we provide an overview of the data used in this paper, highlighting aspects of the data and reductions that are most relevant.  
A brief summary of the spectrographs and representative settings are listed in Table~\ref{observation}.

\begin{figure*}[!t]
\centering
\includegraphics[trim={3.4cm 3.7cm 2.6cm 2cm},width=0.78\textwidth]{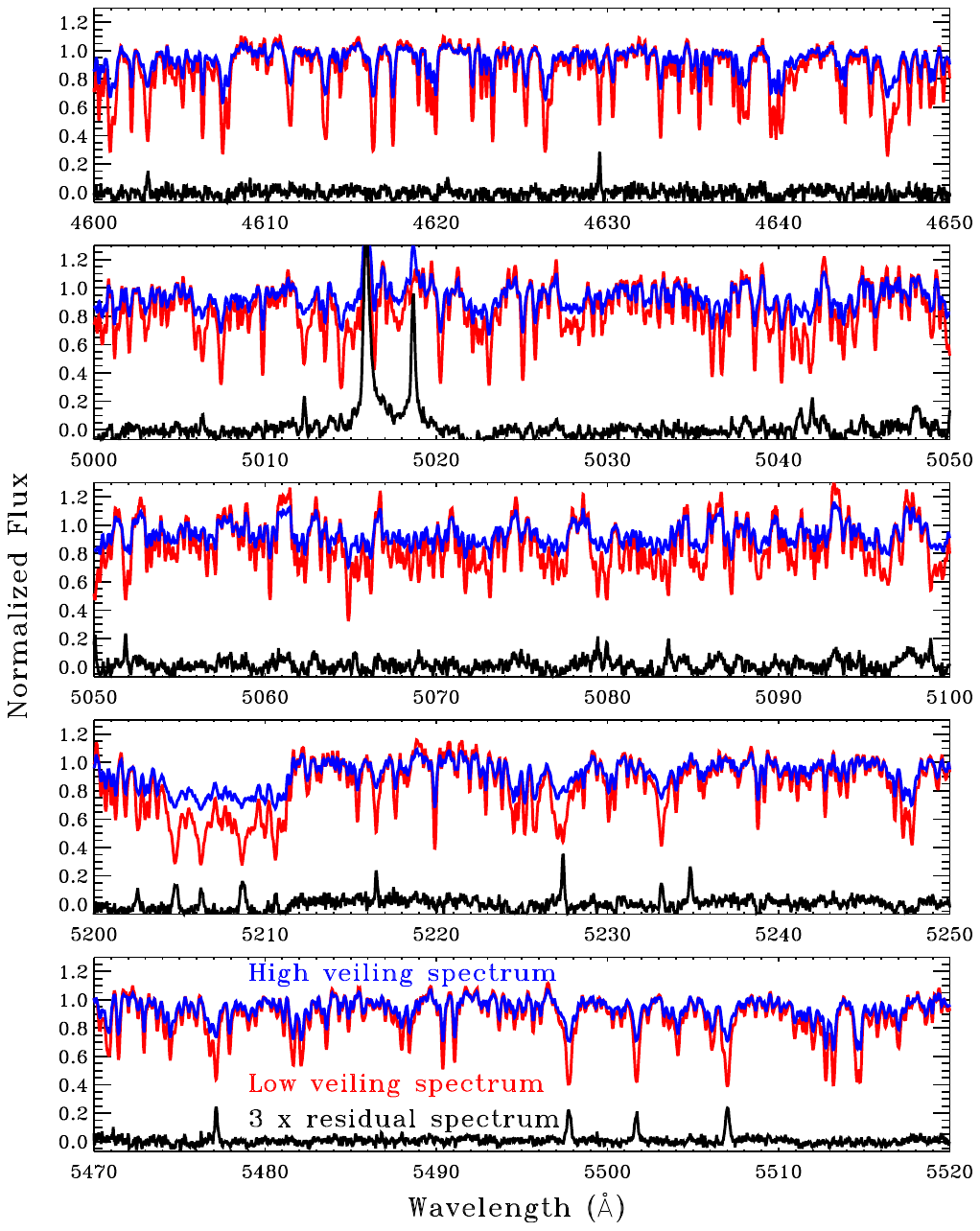}
%\vspace{5mm}
\caption{Coadded ESPaDOnS spectra of TW Hya during visits with strong accretion ($F_{\rm high~veil}$, blue) and weak accretion ($F_{\rm low~veil}$, red) over five wavelength ranges. The residual spectrum ($F_{\rm residual}$, black, scaled by a factor of 3 for visualization) is calculated by veiling the weak accretion spectrum and then subtracting it from the strong accretion spectrum, as $F_{\rm residual}=F_{\rm high~veil} - (F_{\rm low~veil} + r_\lambda)/(1+r_\lambda)$, with $r_\lambda$ as the veiling at wavelength $\lambda$ and with all spectra normalized to the continuum level within the spectral region (normalization shifted slightly in this Figure for visualization).  The residual spectrum reveals emission lines that would be undetectable in any single spectrum, affecting the spectrum by filling in photospheric absorption lines \citep[see examples in][]{gahm08,dodin12}.   Many of the narrow emission lines in the residual spectrum are \ion{Fe}{1} \citep[e.g.][]{hamann92,beristain98}, although line identification is beyond the scope of this paper.  Appendix A describes how the strong, weak, and residual spectra are calculated.}
\label{fig:weaklines}
\end{figure*}

\subsection{High Resolution Spectra}

The analysis for this survey is built from a foundation of 284 spectra obtained with the Echelle SpectroPolarimetric Device for the Observation of Stars (ESPaDOnS) at the Canada-France-Hawaii Telescope (CFHT). 
We also downloaded reduced science spectra of TW Hya obtained with the Fiber-fed Extended Range Optical Spectrograph \citep[FEROS,][]{Kaufer1999} at MPG/ESO-2.2m telescope, the High Accuracy Radial velocity Planet Searcher \citep[HARPS,][]{Mayor2003} at the ESO La Silla 3.6m telescope, in addition to many spectra obtained with other instruments.

The ESPaDOnS spectra were reduced by the automatic data reduction pipeline Libre-ESpRIT \citep{Donati1997}, with some published by \citet{Donati2011}.  In the ESPaDOnS spectra, an additional post-reduction step eliminated noisy regions near order edges to improve the S/N.  The individual orders were backed out from the 1D spectra and then recombined using the hrs\_merge IDL routine, with final spectra that are oversampled at $\sim 0.01$ \AA\ (a factor of a few smaller than the actual pixel scale for ESPaDOnS).  
Most MIKE spectra were obtained and published by \citet{Dupree2012} and \citet{dupree14}, with detailed descriptions of the reductions and calibrations. 
Most HARPS, UVES, and FEROS spectra were automatically reduced and then downloaded from the ESO archive\footnote{https://www.eso.org/sci/facilities/lasilla/instruments/feros/tools/DRS.html}.  
Of the FEROS spectra, 54 spectra were reduced and published by \citet{Alencar2002} and \citet{Batalha2002}.  Most Keck/HIRES spectra were obtained from the Keck Observatory Archive, with automated reductions using the MAKEE\footnote{see https://sites.astro.caltech.edu/\~tb/makee/} pipeline.  The Keck spectra obtained on 2008-01-23 and 2008-05-23 were reduced using a custom-written code in IDL.  The ESPRESSO data were obtained as part of the PENELLOPE program, with reduction described in \citet{manara21}.  

We also obtained high-resolution spectra with CHIRON, a bench-mounted, fiber-fed, cross-dispersed echelle spectrograph \citep{tokovinin13} on the 1.5~m telescope at the Cerro Tololo Inter-American Observatory (CTIO) and is part of
the Small and Moderate Aperture Research Telescope System (SMARTS).
The CHIRON data were taken in ``fiber mode", for a fiber with diameter of 2.7~arcsec on the sky, with $4\times4$ on-chip binning yielding
a resolution $\lambda/\delta\lambda\sim27,800$. Wavelength coverage is
complete from 4080~\AA\ through 8262~\AA\ in 70 orders, with incomplete
coverage to 8900~\AA\ due to interorder gaps between the last five orders.
The data were reduced using a pipeline coded in IDL\footnote{http://www.astro.sunysb.edu/fwalter/SMARTS/CHIRON/ch\_reduce.pdf}, with cosmic ray removal using the L.A.Cosmic
algorithm \citep{vandokkum01}.  

No telluric correction is performed on any high-resolution spectrum.  All measurements in this paper from the high-resolution spectra are made after normalizing the continuum, including corrections for blaze functions.

This paper uses almost all high-resolution spectra that we could find that cover $<6000$ \AA\ and had available, high-quality reductions.  About ten high-resolution optical spectra are excluded from this paper because veiling measurements from independent diagnostics have large standard deviations; several of these excluded spectra were obtained with other instruments not included in this paper.

No selection was made for adverse weather conditions.  Several spectrographs, including ESPaDOnS (in polarimetry mode) and CHIRON, do not have sky fibers, so sky subtraction is not possible.  For bright targets, night-sky emission is generally negligible
apart from narrow [\ion{O}{1}] and Na~D lines and some OH airglow lines at
longer wavelengths.  However, a bright moon and thin clouds can introduce artificial veiling with a blue color.  The CHIRON spectra are excluded from our analysis of [\ion{O}{1}] emission because the spectral resolution is low enough that telluric emission blends easily with the narrow line from TW Hya.

Some high-resolution spectra saturates in H$\alpha$, especially affecting UVES and CHIRON spectra.  Our H$\alpha$ analysis includes only ESPaDOnS, ESPRESSO, and some HARPS and FEROS observations.  Only 10\% of HARPS and FEROS spectra are excluded due to saturation, so any introduction of bias is minimal.  We confirmed using the 2D images that the H$\alpha$ profiles analyzed in this paper are not affected by saturation or linearity.  The only exceptions, where the profile was used but not checked, are old FEROS spectra, which have been previously published and where the line profiles look reasonable by eye.  The exclusion of those old FEROS spectra would not affect the results.

\begin{figure*}[!ht]
\centering
\includegraphics[trim={2.2cm 2.75cm 3.5cm 13cm},width=0.48\textwidth]{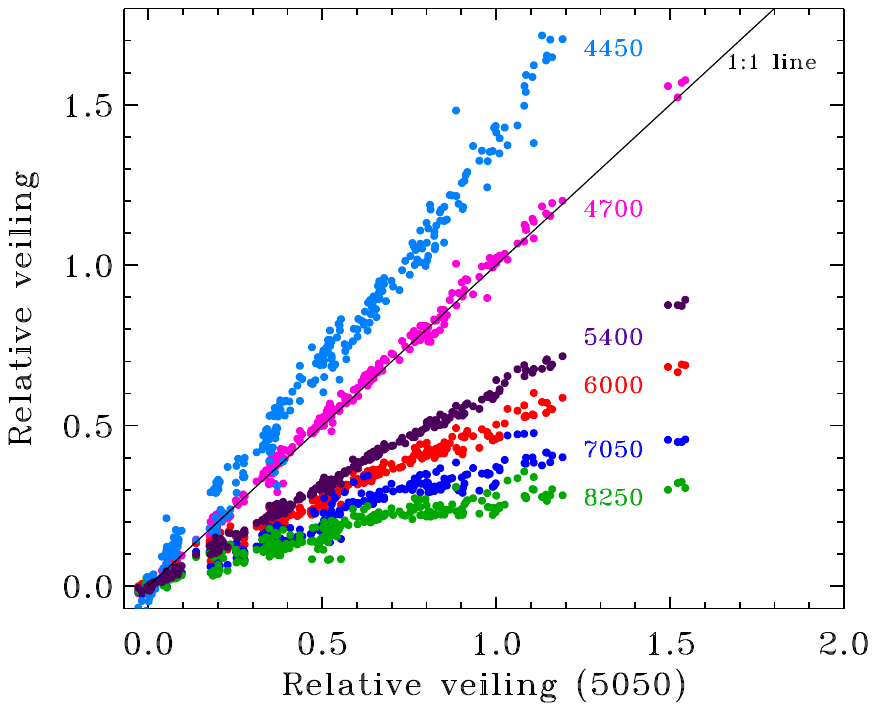}
\includegraphics[trim={2.2cm 2.75cm 3.5cm 13cm},width=0.48\textwidth]{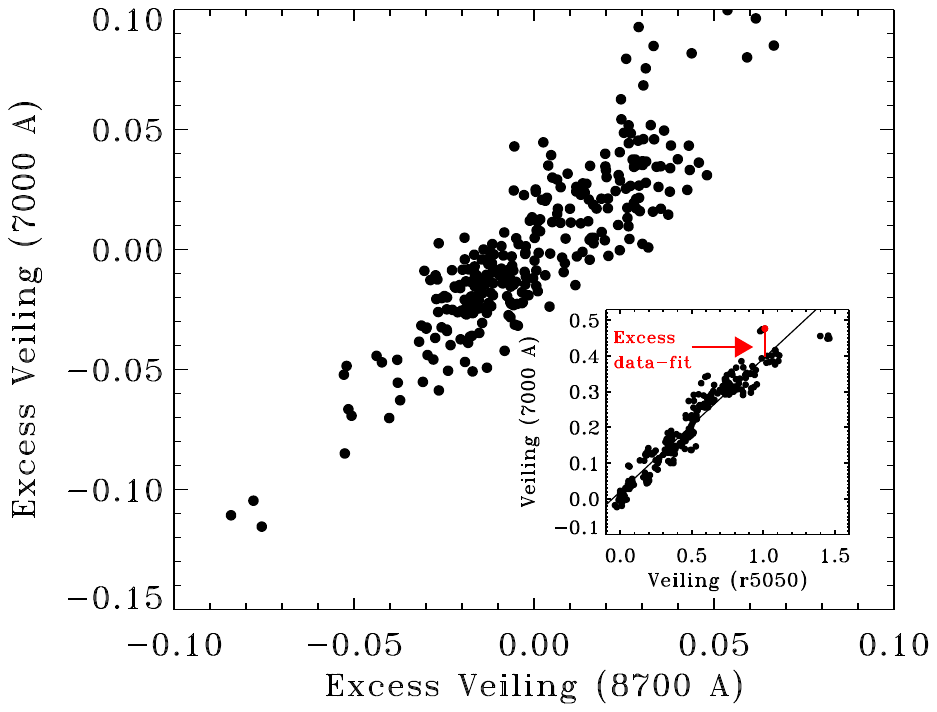}
\caption{{\it Left:} The relative veiling (veiling relative to the low-veiling spectrum of TW Hya) in six wavelength regions (y-axis) compared to the veiling from 5000-5100 \AA\ (x-axis), as measured from ESPaDOnS spectra.  Each veiling measurement plotted here is the average of 4--8 different 25 \AA\ spectral regions around the labeled wavelength.  The line shows a 1:1 relationship and not a fit. The veiling at bluer wavelengths is generally higher than the veiling at red wavelengths, with exceptions. The veiling at 4700 \AA\ is similar to that at 5050 \AA\ because of similar photospheric fluxes.
{\it Right:} Correlated differences between measured and expected veiling at long wavelengths.  The inset shows the correlation between veiling at 7000 \AA\ and 5000-5100 \AA. The excess veiling is calculated by fitting a line to the $r_{\lambda}$-$r_{5050}$ relationship (as seen on the left panel) and then subtracting each point from that line.  The main plot shows this excess veiling at 7000 \AA\ versus that at 8700 \AA.  When the veiling at 7000 \AA\ is higher than expected, the veiling at 8700 \AA\ is also higher than expected. 
The correlation between excess veiling at 7000 and 8700 \AA\ demonstrates that the scatter in the correlations between veiling measurements is real and not due to signal-to-noise or other statistical uncertainties.}
\label{fig:veilcomp}
\end{figure*}

\subsection{Balmer Continuum Spectra}

Accretion rates are most accurately measured from flux-calibrated spectra that cover the Balmer jump at $\sim 3646$ \AA\ \citep[see review by][]{hartmann16}.  Our Balmer continuum measurements of TW Hya are listed 
in Table~\ref{observation} (see also Table~\ref{tab:lowres}).  The STIS data were obtained from the archive in a fully reduced and flux-calibrated format.  The X-Shooter data were reduced following procedures described in \citet{Alcala2014}.  The DBSP, LRIS, and SNIFS data were reduced with custom-build routines in IDL, following \citet{herczeg14} and \citet{guo18}.  The ground-based spectra have fluxes calibrated to $\sim 10$\% with a contemporaneous (within $\sim 1$ hr) spectrum of the spectrophotometric standard LTT 3864, obtained at similar airmass.  
\begin{figure*}[!t]
\centering
\includegraphics[trim=25mm 30mm 20mm 30mm,width=0.49\textwidth]{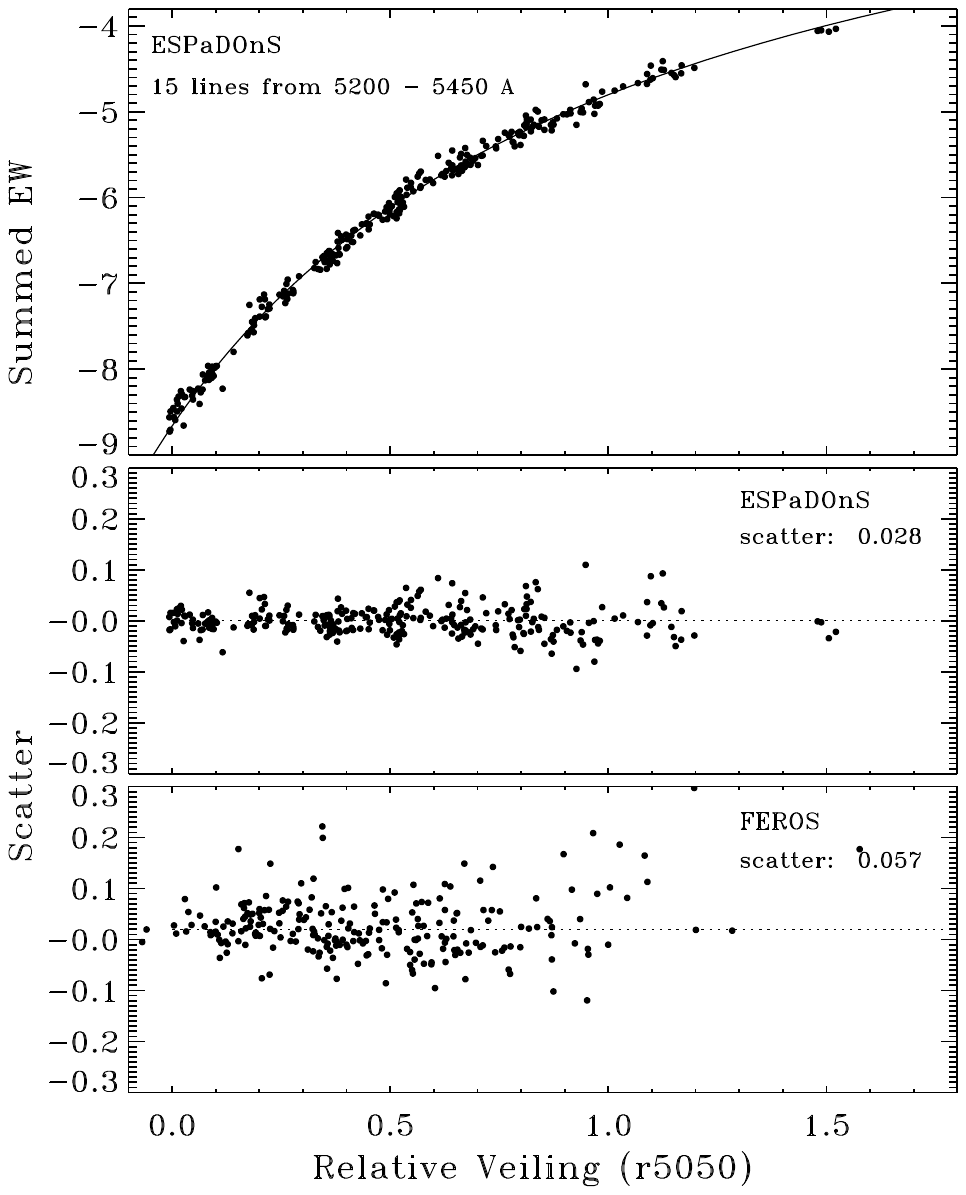}
\includegraphics[trim=25mm 30mm 20mm 30mm,width=0.49\textwidth]{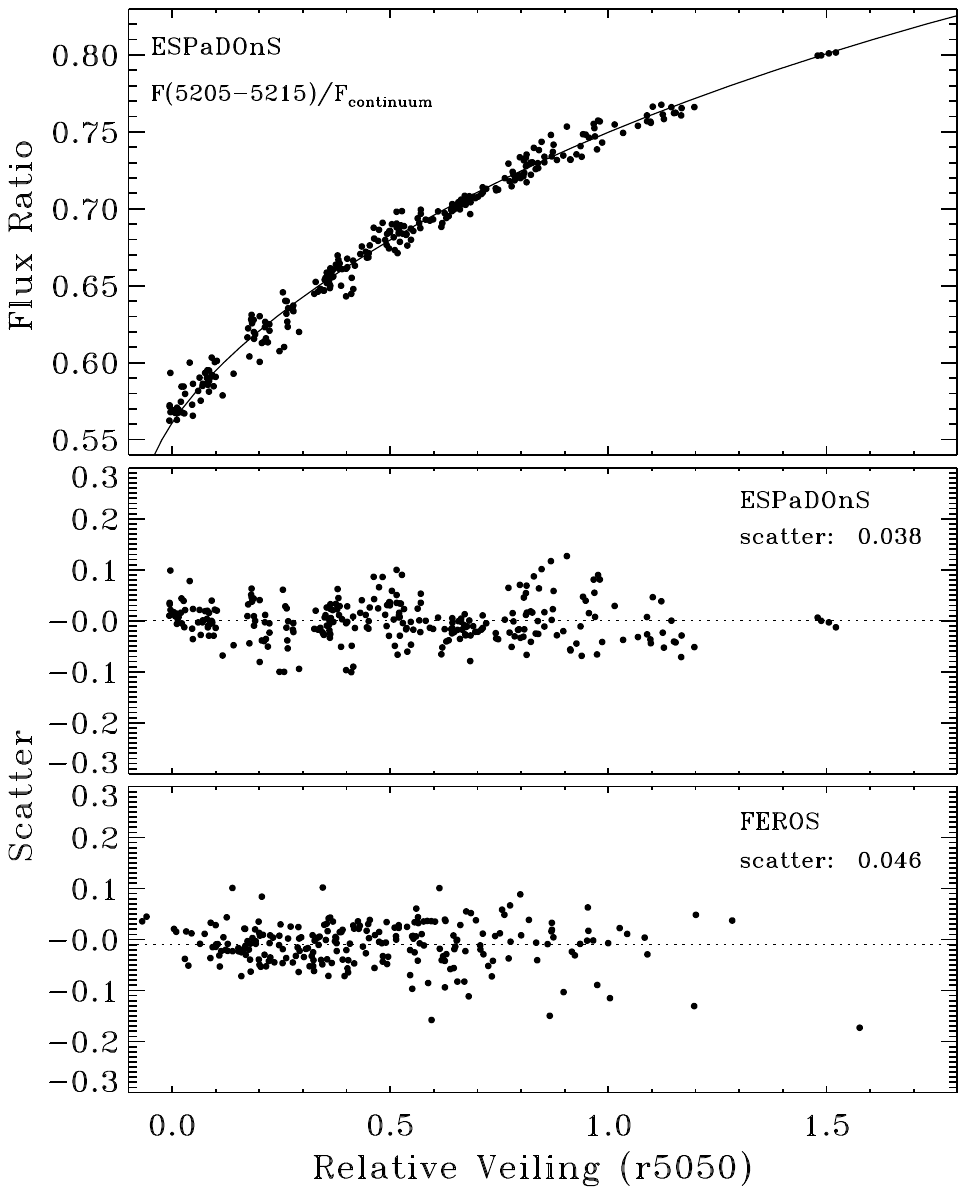}
\includegraphics[trim=20mm 125mm 20mm 20mm,width=0.59\textwidth]{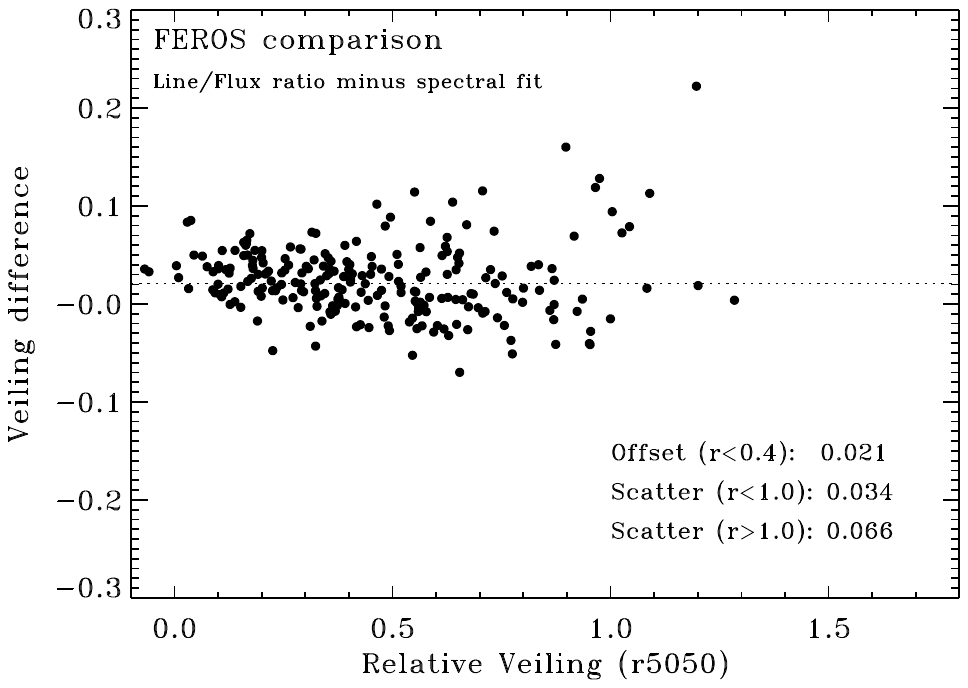}
\caption{{\it Top Left:} The relationship between the equivalent width of coadded lines (one set shown here for lines from 5250--5400 \AA) versus veiling, established from ESPaDOnS spectra and with residuals in ESPaDOnS and FEROS spectra shown below.  {\it Top Right:}  Similar plots as on the left for the flux ratio for a spectral dip around 5205--5215 \AA\ compared with a continuum region.  In both cases, we calculate a best-fit relationship between equivalent width (or flux ratios) and veiling with ESPaDOnS.  We then apply those relationships to FEROS spectra.  The bottom panels on the left and right show the scatter between the veiling calculated from these relationships and the veiling measured from comparing the FEROS spectrum to a low-veiling FEROS template.  
 {\it Bottom:}  The final comparison of veiling obtained from the combination of line equivalent widths and flux ratios to the veiling from a low-veiling FEROS template.  All FEROS spectra are shifted by $\sim 0.02$ to place them on the same scale as the ESPaDOnS veiling measuremenets. 
   The scatter of $0.013+0.045 \times r5050$  (with a minimum error of 0.02) is applied as an uncertainty to all spectra where veiling is measured from line equivalent widths and flux ratios.}
\label{fig:eqwtoveil}
\end{figure*}

\subsection{Spectra for photospheric templates}
Measurements of the excess accretion flux require the subtraction of a photosphere.  The primary spectral template used for this paper is TWA 25, a young star in the same association as TW Hya and with a similar spectral type of ~M0.5 \citep{Herczeg2014}.  Our default TWA 25 spectrum is obtained by stacking 18 ESPaDOnS spectra from 2016--2017 as part of Programs 16AP18 and 17AF95 (PI Donati).  For HARPS spectra of TW Hya, we use as our template HARPS spectra of TWA 25, obtained in 2005--2006 in Programs 076.C-0010 and 074.C-0037.  We also use ESPaDOnS spectra of TAP 45 (K6), V819 Tau (K8), and LkCa 7 (M1.2), each selected for spectral type and relatively narrow photospheric lines.

For low-resolution spectra, we measure the accretion luminosity of TW Hya after subtracting a flux-calibrated low-resolution spectrum of TWA 25 that was obtained with Palomar/DBSP \citep{herczeg14}.  A gap from 5500--6300 \AA\ and emission at $>8700$ \AA\ is filled in with a flux-calibrated Keck/LRIS spectrum of TWA 14.

\section{Measuring the veiling of TW Hya}
\label{sec_veiling}

The optical spectrum of TW Hya consists of photospheric emission combined with line and continuum emission produced by the accretion flow and shock.  Accretion processes produce strong H and He emission lines, H recombination emission, and an H$^-$ continuum, as described and modeled by \citet{Calvet1998}.  The accretion continuum is blue while the photosphere of a young low-mass star is red, so the  ratio of accretion to photospheric flux decreases to longer wavelengths \citep[e.g.][]{basri90,Johns1995,Dupree2012}. 
This ratio of accretion flux $F_{\rm acc,\lambda}$ to 
photospheric flux $F_{\rm phot, \lambda}$ at a wavelength $\lambda$ 
is defined as the veiling, 
$r_\lambda=F_{\rm acc, \lambda}/F_{\rm phot, \lambda}$.  Veiling is measured by comparing the depth of photospheric features in the accreting spectrum to the depth in the spectrum of a photospheric template \citep[e.g.][]{frasca17}.
In \S 4.3 (see also Figure~\ref{fig:veilspec}), a properly scaled accretion continuum will be calculated from our measurements of $r_\lambda$ and a photospheric template.

Traditionally the veiling is measured by comparing the spectrum to another star, chosen to be a low-accretion analog of the target star.  Here we use a two-step process, first calculating the {\it relative} veiling compared to weak accretion epochs of TW Hya and then calibrating those veiling measurements to a zero point.  
  We take three different and mostly independent approaches to measure the relative veiling:  (1) best fits to spectra over short wavelength segments after adding a flat continuum to weak veiling spectra of TW Hya; (2) equivalent widths of strong lines, and (3) spectral indices obtained by comparing regions with weak fluxes to nearby regions.  The analyses are built on the 284 ESPaDOnS spectra, which have high signal-to-noise, consistent spectral resolution, and consistent, high-quality reductions, and are then tailored to spectra from other instruments.  The final veiling measurements used for subsequent analysis are for the 5000-5100 \AA\ spectral region.

\subsection{Methodology for Relative Veiling Measurements}

In this section, we develop methods to measure veiling, relative to a weak accretion spectrum of TW Hya.  We first measure veiling by fitting spectra with a featureless continuum added to the weak accretion spectrum (\S 4.2.1), which works well for fiber-fed spectrographs.  We subsequently measure veiling from stacked lines (\S 4.2.2) and spectral indices (\S 4.2.3), using ESPaDOnS to establish relationships to the veiling measurements and feature depths.
 
The weak accretion spectrum is obtained by summing the 15 ESPaDOnS spectra of TW Hya with the lowest veiling (as measured in \S 4.2.1) and high S/N.  We also create a high veiling template from the strongest 11 spectra.  Figure~\ref{fig:weaklines} shows the comparison between the low and high veiling spectra (for further details on these spectra, see Appendix A).
For other fiber-fed spectra, we also obtain a weak accretion spectrum for the specific instrument and then used stacked lines and spectral indices to scale them to the weak accretion ESPaDOnS spectrum.

\subsubsection{Veiling measurements from spectral regions}
\label{sec:veiling_measurements_direct}

Veiling in the ESPaDonS, HARPS, FEROS, and CHIRON spectra is measured relative to the weak accretion spectrum of TW Hya, obtained with the same instrument in 25 \AA\ intervals across the full observed wavelength region.  This set of values is then converted to a final veiling from 5000--5100~\AA, $r_{5050}$, adopted for the remainder of the analysis.  This wavelength range is selected because it is blue enough for a wide range in veilings, is covered by and well separated from dichroics in low-resolution spectra, and is red enough to be covered by most high-resolution spectra.

To measure the veiling in each 25 \AA\ interval, the weak accretion template is normalized by the median flux in that segment.  A flat featureless emission spectrum is then added to the weak accretion spectrum, and the combined spectrum is renormalized by dividing by $(1+r_\lambda)$.  The best-fit veiling is obtained by minimizing $\chi^2$ in the difference between the science and veiled template spectrum (see also Figure~\ref{fig:weaklines}).

Veiling measurements from all spectral regions are tightly correlated with the veiling from 5000-5100 \AA.  Six examples are provided in Figure 2.  We obtain final veiling measurements for the 5000-5100 \AA\ region, $r_{5050}$, by combining the highest quality
 intervals across the full spectrum, as follows.
First an average veiling, $\bar{r}_{5050}$, is obtained by averaging the veiling from the four intervals between 5000-5100 \AA.  Relationships between $\bar{r}_{5050}$ and $r_\lambda$ (the set of veiling in 25 \AA\ intervals, as in Figure~\ref{fig:veilcomp}) are then fit with a second order polynomial (not shown in the Figure).  These relationships are then used to convert the set of $r_\lambda$ to a set of veilings $r^\prime_{5050}$, now all on the same scale (veiling between 5000-5100 \AA).

For each spectrum, the final veiling measurement, $r_{5050}$, is finally obtained by taking the median $r^\prime_{5050}$ from 80 distinct wavelength intervals.  These wavelength intervals are selected because the scatter between the measured and the preliminary $r_{5050}$ is less than 0.2.  These regions are all located between 4350--6650 \AA.  Regions at longer wavelengths have large scatter, either because of  temporal changes in visual spot coverage or temperature or in the shape of the accretion continuum.
 Comparisons between independent combinations of $r_\lambda^{\prime}$ are consistent with a standard deviation of 0.003, which is adopted as the precision in veiling estimates from ESPaDOnS spectra. 

For the individual veiling measurements, many spectral regions are identified as unreliable and are not used.  Telluric absorption lines, strong emission lines and deep, gravity-sensitive lines are straightforward to identify and avoid.  The region from 6525--6605 \AA\ is excluded from the fit because that entire region is contaminated by H$\alpha$  emission; a similar region is avoided around H$\beta$ and H$\gamma$.  The region at $<4150$ \AA\ is ignored because of low S/N and veilings that are often too high to be accurately measured with our automated method.  Regions with weak emission lines\footnote{These emission lines fill in absorption lines and are only detectable after subtracting off a template.  These lines are more easily detected in heavily-veiled spectra \citep[e.g.][]{gahm08} but the process has been demonstrated with models to also be important for other accretors \citep{dodin12}.
These lines are not identified here but are likely \ion{Fe}{1} and \ion{Fe}{2} lines \citep[see, e.g.,][]{hamann92,beristain98,stempels03}.} 
are identified and avoided by visually comparing the strong accretion and weak accretion templates.
The masked regions and the 2\% of pixels with the largest differences  between the template and science spectrum are excluded from the $\chi^2$ calculation.
All measurements are made after applying small wavelength adjustments ($\lesssim0.5$ \kms\ for ESPaDOnS) from line centroids, as measured in  \S~\ref{sec:eqw}.

\begin{figure*}[!th]
\centering
\includegraphics[width=0.99\textwidth]{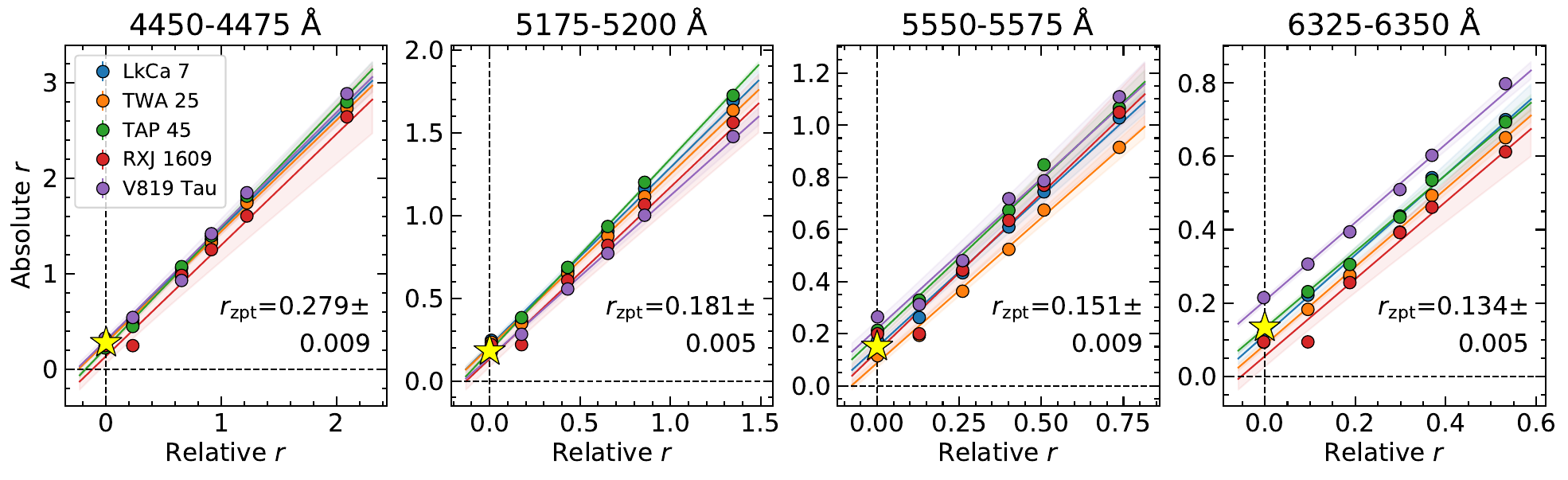}
\caption{Examples of the correlation between absolute veiling measured from WTTS templates and relative veiling measured from weak-accreting spectra of TW Hya itself. Colored dots indicate the measured $r_\mathrm{abs}$ and $r_\mathrm{rel}$ in each sextile with different colors representing different template stars. The solid line and shaded region in the same color as dots is the best-fit linear relation between $r_\mathrm{abs}$ and $r_\mathrm{rel}$ and its 1-sigma error for each template. The yellow stars at $r_\mathrm{rel} = 0$ is the best-fit $r_\mathrm{zpt}$ determined as the inverse-variance weighted mean of the y-intercepts. The $r_\mathrm{zpt}$ measured from different templates are consistent with each other. }
\label{zpt_linear_fit}
\end{figure*}

\begin{figure}[!th]
\centering
\includegraphics[width=0.99\columnwidth]{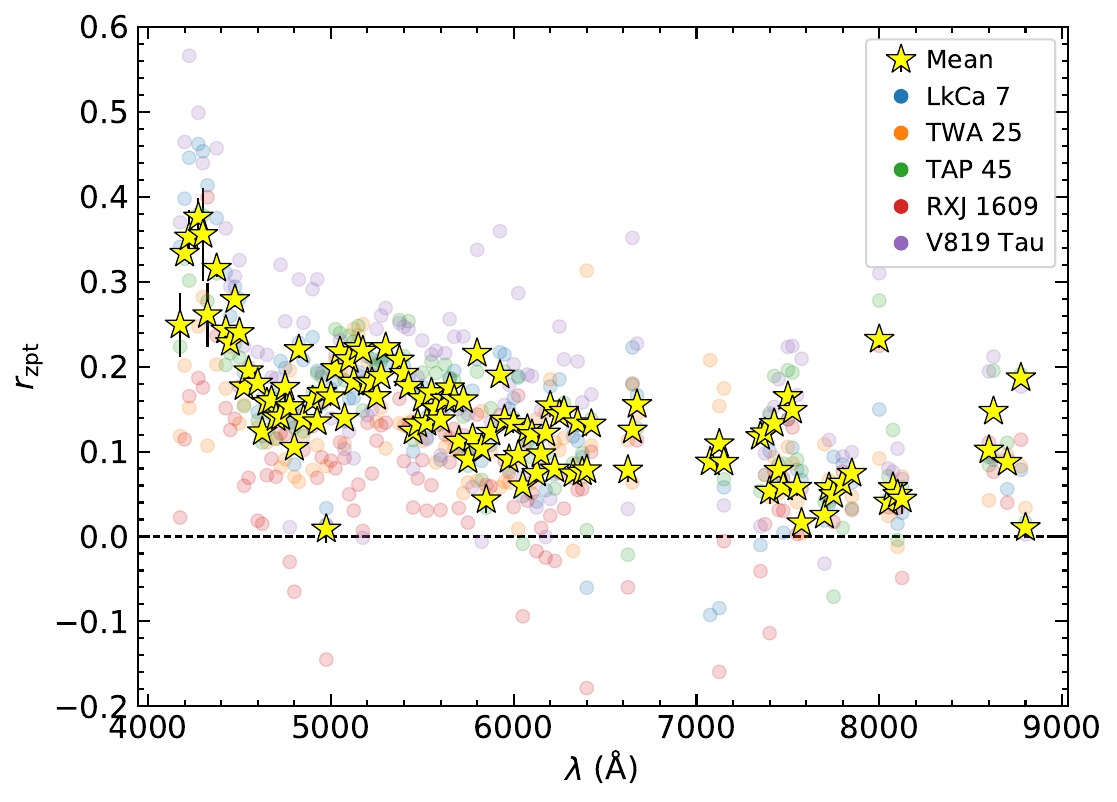}
\caption{The zero-point of relative veiling as a function of wavelength. Yellow stars are the final $r_\mathrm{zpt}$ determined by averaging the individually measured $r_\mathrm{zpt}$ from different stellar templates in colored dots in the background. The shape resembles the relation between relative veiling and wavelength.}
\label{zpt_wave}
\end{figure}

\begin{table*}[!t]
\centering
\caption{Methods and errors}
\begin{tabular}{l|c|cc|cc|ccr}
 & Direct & \multicolumn{2}{c}{Equivalent width} & \multicolumn{2}{c}{Flux Ratios} & \multicolumn{3}{c}{Adopted }  \\
 Instrument  & Scatter & Offset$^a$ & Scatter & Offset$^a$ & Scatter & Relative Error & Random Error & Method\\
\hline
ESPaDOnS & 0.003 & -0.002 & 0.021 & 0.000 & 0.024   & 0.02$^b$ & 0.003 & Direct \\
FEROS    & 0.006 &  0.002 & 0.043 & 0.040 & 0.031 & 0.02 & 0.006 & Direct, shifted\\
HARPS    & 0.003 &  0.068 & 0.026 & 0.099 & 0.019  &  0.02 & 0.003 & Direct, shifted\\
CHIRON   & 0.011 & 0.128 &  0.065 & 0.076 & 0.073 & 0.04 &  0.011 & Direct, shifted\\
\hline
ESPRESSO & -- &--  &--  &--  &--  & -- & 0.02 & EW+Ratios\\
MIKE     & -- & -- &--  & -- &--  & -- & 0.03 & EW \\
UVES     & -- &--  &--  &--  &--  & -- & 0.02 & EW+Ratios\\
HIRES    & -- &--  & -- &--  &--  & -- & 0.02 & EW+Ratios\\
2coude   & -- &--  &--  &--  &--  & -- & 0.03 & EW\\
\hline
\multicolumn{9}{l}{$^a$Difference between veiling from line ratios/equivalent width and veiling measured from low-resolution template}\\
\multicolumn{9}{l}{$^b$Overall offset for all data from zero point analysis}\\
\end{tabular}
\label{tab:errors}
\end{table*}

\subsubsection{Line equivalent widths as veiling measurements}\label{sec:eqw}

As veiling increases, the equivalent width of photospheric absorption lines decreases.  In this subsection, we measure veiling from the depth of coadded photospheric absorption lines\footnote{The procedure and results should be the same by treating each line individually, but results were more robust when first coadding sets of lines.}, following the process in the left panels of Figure~\ref{fig:eqwtoveil}.

We first identify $\sim 260$ lines that are between 4500--6500 \AA\ and are isolated enough to have single-peak absorption profiles.  We then measure equivalent widths for each of the 260 lines from fits with Gaussian profiles.  From this analysis, we select $82$ lines (see Appendix B) that have equivalent widths in ESPaDOnS spectra greater than $0.08$ \AA\ in a median accretion spectrum and that are well correlated with veiling in ESPaDOnS spectra.   These selected lines are then separated by wavelength into 5--6 distinct groups, each with 8-15 lines that are normalized and coadded in velocity space; the exact lines and groupings depend on the wavelength coverage of the instrument.  The equivalent widths for each set of coadded lines are measured by fitting Gaussian profiles to the absorption spectrum.  

The set of equivalent widths from coadded lines is converted to veiling at 5000-5100 \AA\ by fitting a third degree polynomial to the relationship between veiling and the inverse of equivalent width, as measured from ESPaDOnS spectra (Figure~\ref{fig:eqwtoveil}).  
The final veiling from equivalent widths, $r^{\rm EW}_{5050}$, is then measured from the average of the set of 5-6 coadded equivalent widths (see Appendix B).  The number of line sets and the number and categorization of lines into those sets is tailored to the spectral coverage and spectral resolution of the science question.  For CHIRON spectra, some lines that are used with ESPaDOnS are excluded because the lower spectral resolution makes them harder to measure.  The polynomial fits are recalculated for a set of lines that are easily measured with CHIRON and after convolving the ESPaDOnS spectra to the resolution of CHIRON.

 When veiling is measured against a template, the veiling depends on the line, as demonstrated in Figure~\ref{fig:weaklines}.  \citet{rei18} found that higher veiling values are measured from stronger photospheric lines.  Since our equivalent widths are converted to relative veiling through correlations rather than directly comparing line depths to non-accreting templates, they are robust to effects introduced by line-dependent veiling.
 
\begin{figure*}[!ht]
\centering
\includegraphics[trim={4cm 2.85cm 4cm 15cm},width=0.48\textwidth]{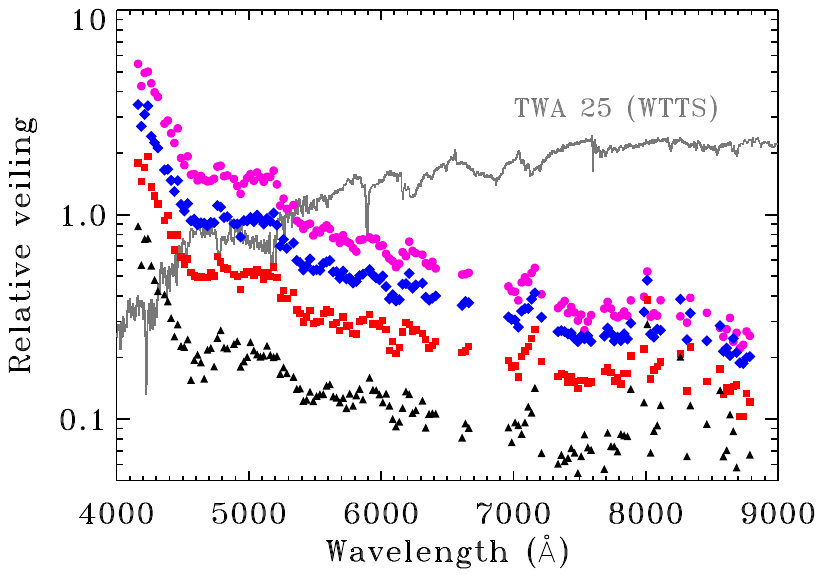}
\includegraphics[trim={4cm 2.85cm 4cm 15cm},width=0.48\textwidth]{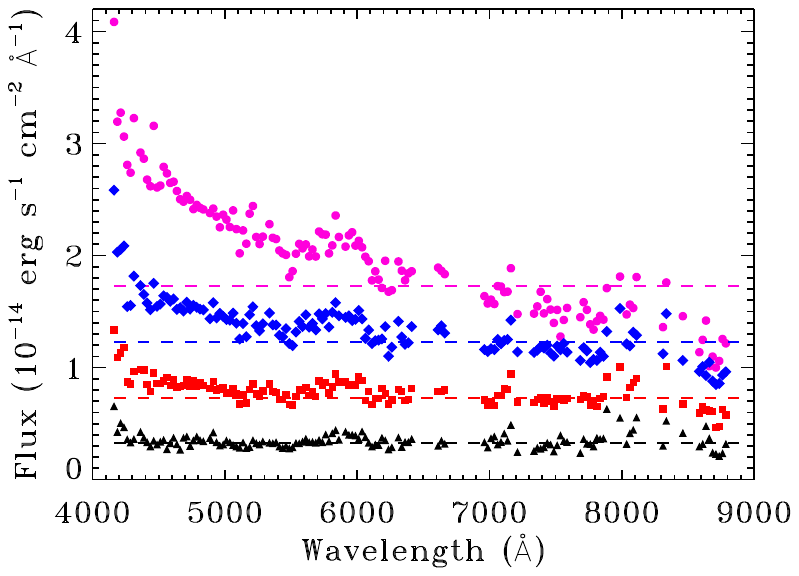}
\caption{Veiling spectra for strong (pink circles, average relative veiling $r_{5050}=1.52$), above average (blue diamonds, $r_{5050}=0.95$), below average (red squares, $r_{5050}=0.52$), and weak (black triangles, $r_{5050}=0.21$) veiling epochs measured in ESPaDOnS spectra.  The left panel shows the relative veiling measurement, with features in the veiling spectrum caused by the shape of the photospheric template, TWA 25.  The right panel shows the veiling spectrum multiplied by TWA 25.  The spectra with moderate accretion are flat from 4000--9000 \AA, while the strongest accretion spectrum has a bluer slope (increasing flux with decreasing wavelength).}
\label{fig:veilspec}
\end{figure*}

\subsubsection{Spectral indices as veiling measurements}

Across the spectrum, some regions of strong line blends have fluxes that are much lower than nearby regions. The spectral index for these regions (the weak region divided by the strong region, or vice versa) depends on veiling by the accretion continuum.  In this subsection, we follow similar procedures as in \S4.1.2 to measure the flux ratios and then convert the spectral index to veiling, following the right panels of Figure~\ref{fig:eqwtoveil}.

We measure spectral indices in seven relatively narrow regions with faint emission divided by nearby spectral regions with brighter emission (see Appendix B).  In the ESPaDOnS data, seven spectral indices are tightly correlated with veiling.  Six additional regions were excluded because the correlations with veiling had large scatter.  For several instruments, we discard some of the five indices because either the spectral region or background continuum region is located near the edge of an echelle order.  We also avoid TiO bands because they are sensitive to spots and are located at long wavelengths, where weak veiling limits the lever arm and leads to large fractional uncertainties.

The relationship between each spectral index and veiling, $r_{5050}$ for our ESPaDOnS sample, is fit with a fourth degree polynomial.  Each spectral index is then converted into an estimate for veiling, with the median value adopted as the veiling measurement, $r^{\rm rat}_{5050}$.  The relative veiling measurements are only applied for the range in ESPaDOnS veiling.  The flux ratios and polynomial fits are recalculated for each instrument by degrading either the resolution of either the ESPaDOnS spectra or the science spectrum so that they match.

\subsection{Applying relative veiling methods to spectra}

Figure~\ref{fig:eqwtoveil} and  Table~\ref{tab:errors} summarize how the methods and the associated errors are applied to spectra from each instrument.  The veiling measurements from ESPaDOnS spectra form the backbone of our analysis. 
The FEROS, HARPS, and CHIRON spectra all have veiling initially measured relative to low-veiling spectra from the same instrument.  These veilings are then offset to the ESPaDOnS values (calculated only for spectra with $r_{5050}<0.5$) by using the correlations of veiling with equivalent widths and flux indices.  For the slit-based spectra and for ESPRESSO (with only 5 spectra), the veiling measurements are obtained by averaging the veiling estimated from the equivalent widths and spectral indices.  For most MIKE spectra and one FEROS spectrum, the veiling is measured from the equivalent width of coadded photospheric lines, which is the method most robust to uncertainties in the relative wavelength solution.

The analysis of uncertainties is split into the precision of the veiling measurements relative to each other (\S 4.2.1) and calibrating all of the measurements against a pure photosphere (\S 4.2.2).

\subsubsection{Relative errors between measurements}

For ESPaDOnS, the veiling measurements are adopted directly from the set of $\sim 80$ veiling values.  Random combinations of different sets of veilings indicate a precision of 0.003.  Similarly, the precision of FEROS, HARPS, and CHIRON veiling measurements ranges from 0.003--0.006.  Those sets of veiling values are scaled to ESPaDOnS, with a relative error of $\sim 0.02$, by using results from the equivalent width and spectral index analysis for spectra with $r_{5050}<0.4$.

For the slit spectra and ESPRESSO, the combined use of equivalent widths and spectral indices leads to uncertainties of $\sim 0.02$ for $r_{5050}<0.3$ and $\sim 0.05$ for $r_{5050}>1$.  The scatter versus veiling is well approximated by $0.013 + 0.045\times r_{5050}$ and is adopted (with a minimum error of 0.02) as the assessed uncertainty.  These values are calculated from the dispersion in the individual measurements and from the scatter in the comparison between the FEROS values for the direct veiling measurements and the veilings measured from line equivalent widths and spectral indices.

On eight occasions, two different instruments observed TW Hya within the same 3 hr window (Table~\ref{tab:checks}).   
The most significant deviation occurred for two observations separated by 62 minutes on MJD 59310, when the veiling was high.  This large discrepancy may be caused by a rapid decline of the veiling as an accretion burst faded or by artificial differences due to larger errors when the veiling is high.  For other epochs, the average difference in veiling is 0.06.  

This analysis, though limited, supports our error estimates.  A few spectra are likely affected by clouds and the moon and may have uncertainties that are underestimated.

\begin{table}[!t]
\centering
\caption{Cross-instrument comparisons of relative veilings}
\begin{tabular}{llcclc}
\hline\hline
MJD & Inst. 1 & $r$ & $\Delta t$ (min) & Inst. 2 & $r$\\
\hline
54158.382 & FEROS & 0.98 & $<20^a$ & MIKE & 0.95 \\  
54159.114 & FEROS & 0.46 & $<20^b$ & MIKE & 0.47 \\  
59280.173 &  CHIRON & 0.59 & 163 & ESPRESSO & 0.48 \\
59308.039 & ESPRESSO & 0.61 & 131 & CHIRON & 0.61\\ 
59309.135 & CHIRON & 1.24 & 11 & ESPRESSO & 1.39 \\ 
 59310.086 & ESPRESSO & 1.47 & 62 & CHIRON & 1.23\\
 59313.146 & CHIRON & 0.08 & 102 & ESPRESSO & 0.06\\
 59667.139 & CHIRON & 0.71 & 127 & UVES & 0.84\\
       \hline
       \multicolumn{6}{l}{$^a$median of 6 spectra obtained within 20 minutes of FEROS}\\
       \multicolumn{6}{l}{$^b$median of 7 spectra obtained within 20 minutes of FEROS}\\
\end{tabular}
\label{tab:checks}
\end{table}

\subsubsection{Zero-point measurements for veiling} \label{sec:zero}

The primary shortcoming of using TW Hya as its own template is in calibrating the veiling measurements to a zero-point, $r_\mathrm{zpt}$.  Our analysis described above provides veiling measurements relative to a low accretion spectrum of TW Hya.  To convert these relative veiling measurements into an absolute veiling relative to a non-accreting star, the veiling of the low accretion spectrum needs to be calculated through comparisons with non-accreting spectral templates.

\begin{table}[!b]
\centering
\caption{Spectral Templates}
\begin{tabular}{lccc}
Star & SpT$^a$ & $v \sin i^b$ & Ref.$^c$ \\
\hline
TW Hya & M0.5 & 4 & \citet{Donati2011}\\
\hline
TAP 45 (V1076 Tau) & K6 & 7.7 & \citet{nguyen12} \\
V819 Tau & K8 & 9.5 & \citet{donati15}\\
TWA 25 (V1249 Cen) & M0.5 & 11.9 & \citet{nicholson21}\\
LkCa 7 (V1070 Tau) & M1.2 & 14.7 & \citet{nguyen12} \\
%RX J1609.5-3850 & --
%TWA 7 & -- & 4.5 \\
\hline
\multicolumn{3}{l}{$^a$From \citet{Herczeg2014}.} & 
\multicolumn{1}{l}{$^b$km/s}\\
\multicolumn{4}{l}{$^c$Reference for $v \sin i$ measurements.}\\
\end{tabular}
\label{tab:templates}
\end{table}

We measure a wavelength-dependent zero-point by fitting stacked TW Hya spectra with the spectra of  non-accreting stellar templates (Table~\ref{tab:templates}) with spectral types from K6--M1.5 to bracket that of TW Hya.
These non-accreting stars are all young\footnote{Previous analyses had adopted the K6 dwarf star GJ 1172 as a spectral template for TW Hya \citep[e.g.][]{Alencar2002,Dupree2012}, however GJ 1172 in particular has much shallower TiO absorption than TW Hya and does not provide a good fit over the full optical wavelength range.}, selected to match the gravity and chromospheric effects of TW Hya  (see, e.g., discussions of templates in \citealt{Ingleby2013} and \citealt{manara13}).

The spectral fits are similar to those described in  \S\ref{sec:veiling_measurements_direct}, with additional free parameters \vsini\ and radial velocity alongside the veiling. The rotational broadening kernel is adopted from  \citet{Gray2005} and does not include limb-darkening.
The TW Hya spectra used in this fit are mean-stacked (ESPaDOnS) spectra in six equal-sized bins of relative veilings. The absolute veiling ($r_\mathrm{abs}$) measured from this fit are correlated with the mean relative veiling ($r_\mathrm{rel}$) of each sextile measured from weak-accreting TW Hya spectra (Figure \ref{zpt_linear_fit}).

Figure \ref{zpt_wave} shows the relation between $r_\mathrm{zpt}$ and wavelength, which closely resembles the shape of relative veiling as shown in Figure~\ref{fig:veilspec}. The $r_\mathrm{zpt}$ is not correlated with the spectral type of the template.  
The systematic error of $r_\mathrm{zpt}$, estimated as the median RMS of the $r_\mathrm{zpt}$ determined from different templates, is $\simeq 0.02$.

\begin{figure*}[!t]
\centering
\includegraphics[trim={2.5cm 12.9cm 2.5cm 2.3cm},width=0.48\textwidth]{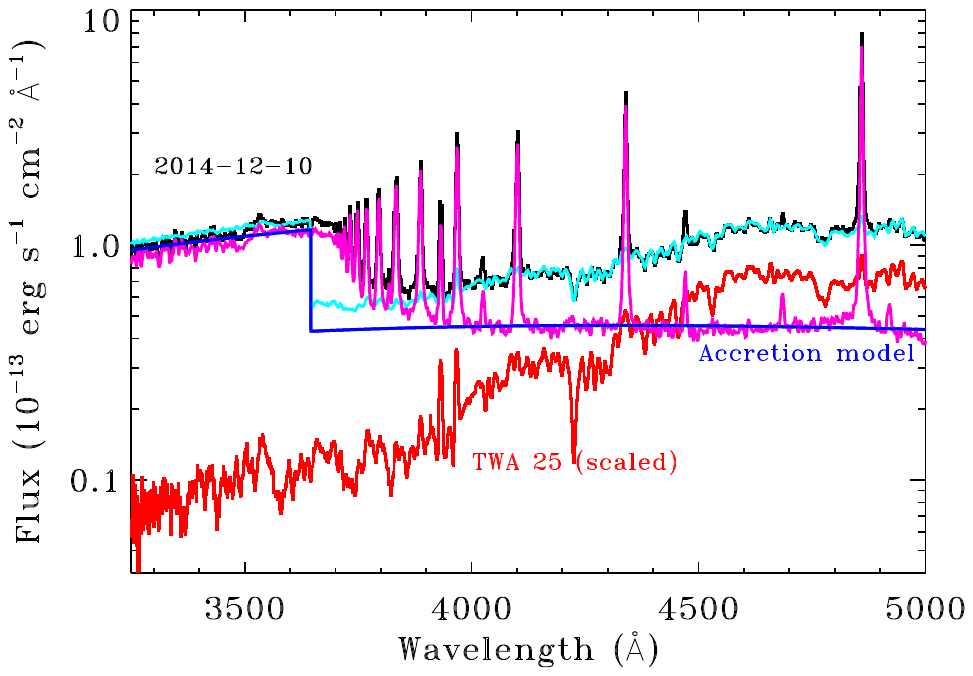}
\includegraphics[trim={2.5cm 12.9cm 2.5cm 2.3cm},width=0.48\textwidth]{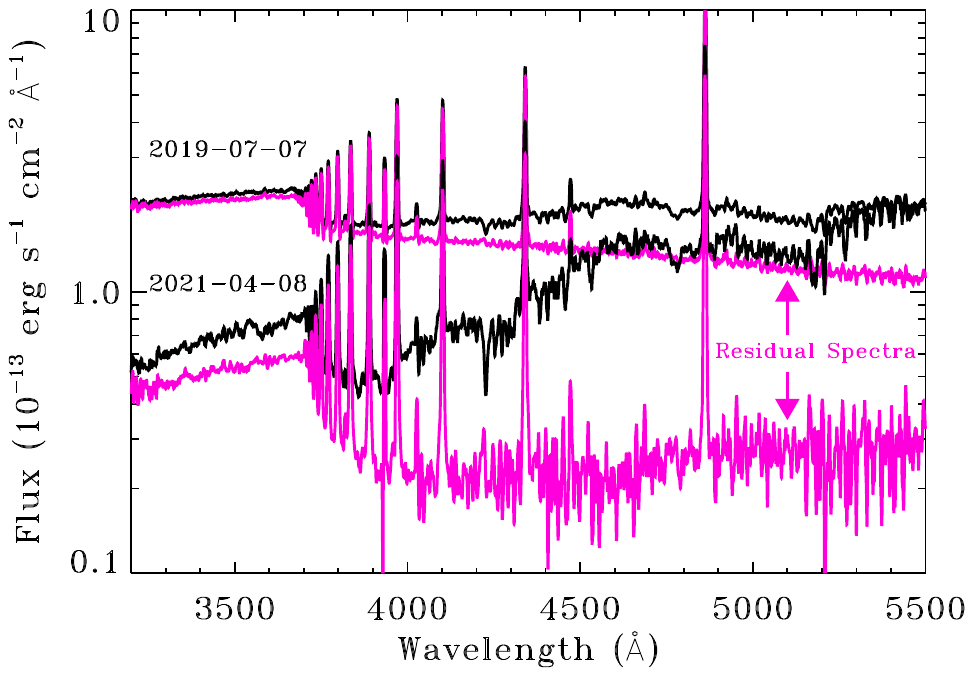}
\caption{The accretion luminosity measured from flux-calibrated spectra of TW Hya.  The left panel shows the spectrum of TW Hya (black), modeled (light blue) as the sum of a stellar photosphere (red) and accretion continuum (blue), with a residual accretion spectrum (pink) calculated  by subtracting the scaled photosphere from the TW Hya spectrum.  The right panel shows the TW Hya spectrum and the residual accretion spectrum on an epoch of strong accretion (2019-07-07) and weak accretion (2021-04-08).  The weak accretion spectra tend to have larger Balmer jumps, while the strong accretion spectra have a bluer slope (increasing flux to shorter wavelengths).}
\label{fig:balmerjump}
\end{figure*}

\begin{table}[!t]
\centering
\caption{Veiling and Accretion Measuremens}
\begin{tabular}{lccccc}
\hline
 MJD & $r_{5050}$ & $\sigma(r_{5050})$ & $L_\odot$ & $M_\odot$ & Inst. \\
 \hline
    53842.20872   &  1.07   &  0.056  &  0.057 & 3.15e-09  &  MIKE\\
    53842.96101  &   0.59   &  0.037  &  0.032 & 1.81e-09  &  MIKE\\
    54225.09375   &  0.28   &  0.003  &  0.017 & 9.38e-10  &  FEROS \\
    59241.30469  &    0.24  &    0.012  & 0.015 & 8.43e-10  &  ESPaDOnS \\
    59309.13672  &   1.41   &  0.018  &  0.074 & 4.10e-09  &  ESPaDOnS\\
    \hline
    \multicolumn{6}{l}{Five dates selected randomly.  Full table is online.}
\\
\end{tabular}
\label{tab:veil}
\end{table}

\subsection{Summary of veiling measurements and the accretion spectrum}

The combined approaches from the analysis above yield veiling measurements for 1169 spectra.  The final veiling values for the 5000-5100 \AA\ region range from 0.13--2.13 with a median of 0.67, after excluding observations obtained within 3 hrs of another observation.

The uncertainties in the veiling are small enough to be negligible relative to other errors.  The relative error in most accretion rate estimates is dominated by uncertainties in how the bolometric correction varies with time and by the change in the photospheric spectrum due to cool photospheric spots.  In principle, an increase (or decrease) in accretion spots (or a change in accretion spot coverage in the visible surface) could also decrease (or increase) the observed photospheric emission, although usually the accretion spots only cover a few percent of the stellar surface \citep[e.g.][]{Calvet1998}.
The zero-point error in veiling of $0.02$ leads to a fractional uncertainty of 10-15\% during periods of the weakest accretion.

The ESPaDOnS spectra yield veiling measurements across the 4150--9000 \AA\ wavelength region, so a veiling at any wavelength is accurately converted to a veiling at any other wavelength, with a relative uncertainty of $\sim 0.01-0.03$.  Figure~\ref{fig:veilspec} shows that the veiling increases to short wavelengths, as expected for hot emission against a cooler photosphere.  Spectral features in the photosphere, including at 5200 \AA\ and TiO 7140 band, are seen as sharp changes in the veiling because the nearly flat accretion continuum is divided by a photosphere with features.

The accretion spectrum is obtained by convolving veiling with the flux-calibrated template spectrum of TWA 25.  For low and modest veiling values ($r<0.5$, corresponding to $0.025$ L$_\odot$, see \S 5.1), the accretion spectrum is consistent with a constant flux across the optical spectral range.  When the veiling is high ($r>1.2$, or $L_\odot\sim0.06$ L$_\odot$), the accretion spectrum becomes stronger at blue wavelengths.  

Figure~\ref{fig:veilcomp} also shows that the excess veiling at red wavelengths is correlated with excess veiling at other red wavelengths.  This {\it excess} veiling is the difference in veiling between the measured veiling at wavelength $\lambda$ and the veiling expected from the final veiling $r_{5050}$ and $\lambda$.  If the scatter in those correlations were a consequence of measurement uncertainties, then the excess veiling at two wavelengths would be uncorrelated.  The correlation between the excess veiling at 7000 and 8700 \AA\ demonstrates that the scatter at red wavelengths is real.  This scatter may be caused by either spots or by changes in the temperature of the accreting gas.

\section{Converting Veiling Measurements to Accretion Rate}
\label{sect_rtomdot}

The primary goal of this paper is to analyze the stability of the accretion rate with time.  In this section, we convert the veiling measurements in \S 4 to accretion rates by first measuring accretion luminosities from
flux-calibrated spectra of TW Hya (\S 5.1), and then find a relationship between the accretion luminosity and the veiling at $5000-5100$ \AA\ (\S 5.2).  Finally, we use that relationship to calculate accretion rates from the sample of high-resolution spectra (\S 5.3).

The accretion rates are measured from broadband, flux-calibrated spectra.  Assuming that all gravitational energy is converted into luminosity and following \citep[]{Gullbring1998}, the accretion rate is calculated  by
\begin{equation}
\label{mdotew}
\dot M_{\rm acc}=(1-\frac{R_*}{R_{\rm in}})^{-1}\frac{L_{\rm acc}R_*}{GM_*} \sim 1.25\frac{L_{\rm acc}R_*}{GM_*} ,
\end{equation}
where $R_*$ and $M_*$ are the stellar radius and mass.  For convention and consistency with other estimates, we adopt R$_{\rm in}\sim5~$R$_*$ as the disk truncation radius \citep[e.g.][]{johnskrull07,Johnstone2014}.  The truncation radius of $\sim 3.5$ R$_*$ meaured by \citet{gravity_garcialopez20} would increase the accretion rate by $12\%$.

\begin{figure*}[!t]
\centering
\includegraphics[trim={2.5cm 13cm 2.5cm 3cm},width=0.48\textwidth]{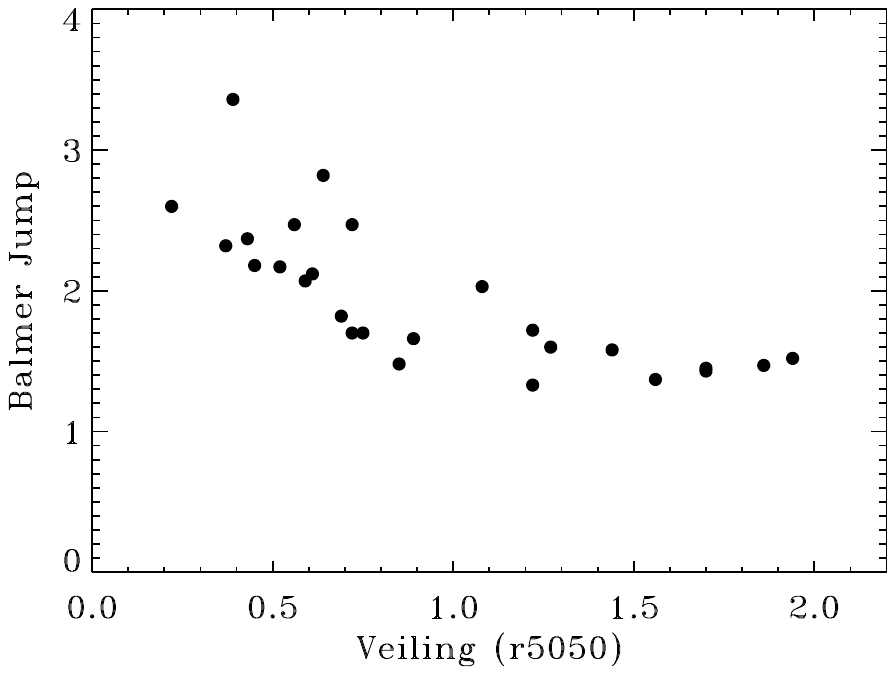}
\includegraphics[trim={2.5cm 13cm 2.5cm 3cm},width=0.48\textwidth]{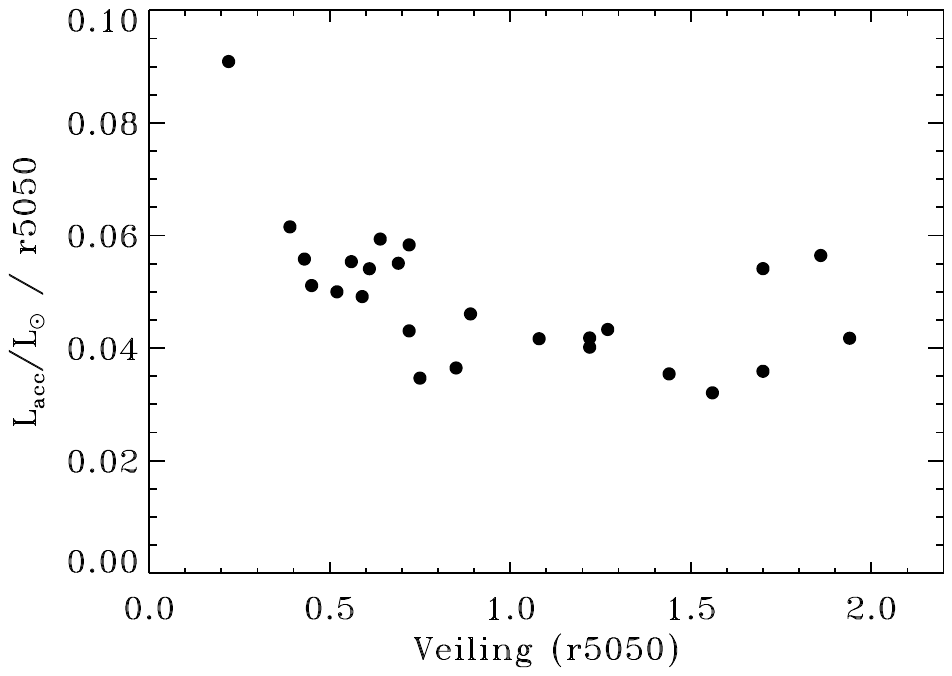}
\caption{Left: The size of the Balmer jump (here defined as the flux ratio of 3600 to 4200 \AA) versus the veiling estimated from low-resolution spectra.  The Balmer jump tends to be higher when the veiling is lower. When the Balmer jump is larger, more emission escapes at short wavelengths, so the bolometric correction is larger.  
Right: The bolometric correction, used to convert the measured accretion flux at 5050 \AA\ to the total accretion luminosity, versus the veiling, as measured from low-resolution spectra.}
\label{fig:bolcor}
\end{figure*}

The flux-calibrated spectra are fit with the combination of an accretion spectrum and a low-resolution spectrum of the photospheric template TWA 25.  The spectra of TW Hya shows excess emission at all optical wavelengths, with an increase shortward of the Balmer jump at 3646 \AA.  The Balmer jumps measured in this paper are the ratio of the excess emission at 3600 \AA\ to the excess emission at 4200 \AA, after subtracting the photospheric template.
 The accretion spectrum across the Balmer jump is modeled with a plane-parallel slab (\citealt{Valenti1993}, as implemented by \citealt{Herczeg2008}), while the Paschen and H$^-$ continua are assumed to be featureless and constant in flux, when possible.  
The best-fit combination of a photosphere and accretion continuum is determined by eye to minimize residuals, with a veiling that is generally accurate to $\sim 0.05$. The spectra obtained at medium or high resolution are convolved to low-resolution to match the spectrum of TWA 25.
 The accretion luminosity for each epoch is calculated by summing the total luminosity in the best-fit model.  
 
 Figure~\ref{fig:balmerjump} shows examples of these fits. The residual after subtracting the photosphere should have no significant absorption feature.  In periods of strong accretion, the residual emission (the accretion continuum, after subtracting the photosphere) declines in flux to longer wavelengths, while epochs with weak or moderate accretion have an accretion continuum consistent with a constant flux.  The acretion continuum on 2019-07-07 is described by $F_{\rm acc}=F_{\rm acc,5000}\times (1 -2.5\times10^{-4} (\lambda-5000)$, consistent with the slope in Figure~\ref{fig:veilspec}. The example epoch with weak accretion, on 2021-04-08, has a slope that is consistent with a constant flux.

 The accretion luminosities depend on the assumptions in the broadband accretion spectrum, especially where the accretion emission is unobserved.
  Table~\ref{tab:mdotcomp} shows the comparison between our accretion rates and literature accretion rates, as measured from the same spectrum.
   Our accretion luminosities for the X-Shooter spectra 
are almost exactly the same (difference of $\sim 3\%$) as the values from \citet{manara14} and \citet{venuti19}, both of which applied an independent model for hydrogen slab emission\footnote{Our bolometric correction for 5000-5100 \AA\ also matches the bolometric correction of a 9000 K blackbody.}.  However, recent accretion  models include multiple accretion shocks with a range in temperatures to better explain the veiling at red wavelengths \citep[e.g.][]{Ingleby2013,robinson19,espaillat22,pittman22}.  The introduction of these components significantly enhances the flux at red wavelengths, with accretion luminosities that are 30--60\% higher in \citet{robinson19} than the accretion luminosities measured here. The veiling of TW Hya in the near-IR is low and much more stable \citep{sousa23}, because either the cooler accretion component or the disk emission varies less than the hotter emission.

 \subsection{Bolometric corrections for high resolution spectra}

 The fits to the flux-calibrated spectrum of TW Hya combine an accretion spectrum and a photospheric spectrum.  In this section, we use those fits to derive a relationship to convert  veiling at 5000-5100 \AA\ to accretion luminosity.
 
 The accretion luminosity is calculated by measuring the accretion spectrum and veiling in 26 flux-calibrated blue spectra of TW Hya (see examples in Figure~\ref{fig:balmerjump}).  
The veiling measured at high-resolution is 20\% lower than the veiling measured from low-resolution spectra, in a limited comparison of one Keck/HIRES and two VLT/UVES spectra.
%\footnote{The VLT/UVES spectra used here extend from 3300--6800 \AA\ but do not have an accurate flux calibration, so they are not listed in Table~\ref{tab:lowres}}. 
The difference in veiling is likely attributed to the definition and normalization of the photospheric continuum level, which should be lower when many lines blend together.

Figure~\ref{fig:bolcor} shows our conversions from flux to luminosity.  The average bolometric correction is $F_{5050}/F_{acc}=(1.33\pm0.12)\times10^{-4}$, where this uncertainty is calculated from the standard deviation between the data points and best fit line.  This bolometric correction is applied to the veiling for TW Hya, as 
\begin{equation}
    L_{\rm acc}=0.050 \times r_{\rm lowres} =  0.062 \times r_{\rm highres}
\end{equation}
with an uncertainty of $\sim 10\%$ that is consistent with typical uncertainties in flux calibration.  The low-resolution calculation is obtained directly from average value on the right panel in Figure~\ref{fig:bolcor}, while the conversion for high resolution (the right side of the equation) adjusts for the $\sim 20\%$ difference between the veiling measured from the high and low-resolution spectra (described above).  This equation now allows us to convert our set of 1169 veiling measurements to an accretion luminosity.

The bolometric correction depends on the shape of the accretion continuum, which may vary with time and accretion rate.  Figure~\ref{fig:balmerjump} shows that at the highest veiling measurements, the shape of the accretion continuum (after subtracting off the photosphere) becomes bluer.  This same result is seen in our fits to the flux-calibrated low-resolution spectra (Figure~\ref{fig:veilcomp}).  The size of the Balmer jump in the accretion continuum is also smaller when the accretion rate is higher, a trend that is consistent with higher temperatures or opacities in the accretion shock  \citep[e.g.][]{Calvet1998}.  At low-modest veilings, when $r_{5050}<1$, the consistency in shape of the accretion continuum indicates that the increase in accretion rate is at a constant temperature, so the spot size likely increases.  At high veilings, the bluer spectrum indicates that the increase in accretion rate corresponds to a higher temperature.

\begin{table}
\centering
\caption{Accretion rates and veiling in low resolution spectra}
\begin{tabular}{lcccccc} 
\hline
Instr. & Date & $r_{low}^a$ & $F_{\rm acc}^b$  & BJ & L$_{acc}^c/L_\odot$ & Ref\\
%& & & 1e-14 & -- & $L_\odot$ & \\
\hline
\hline
%update for distance
STIS & 2000-05-07 & 1.08 & 5.99e-14 & 2.03 & 0.045 & H04\\
STIS & 2002-07-19 & 0.64 & 4.23e-14 & 2.82 & 0.038 & HH08\\
DBSP & 2008-01-18 & 0.52 & 3.27e-14 & 2.17 & 0.026 & HH14\\
DBSP & 2008-01-19 & 0.61 & 4.12e-14 & 2.12 & 0.033 & HH14\\
DBSP & 2008-01-20 & 0.69 & 5.07e-14 & 1.82 & 0.038 & HH14\\
LRIS & 2008-05-28 & 0.39 & 2.43e-14 & 3.36 & 0.024 & H09\\
DBSP & 2008-12-28 & 0.45 & 2.92e-14 & 2.18 & 0.023 & HH14\\
DBSP & 2008-12-29 & 0.56 & 3.48e-14 & 2.47 & 0.031 & HH14\\
DBSP & 2008-12-30 & 0.72 & 4.20e-14 & 1.70 & 0.031 & H14\\
STIS & 2010-01-28 & 1.44 & 7.41e-14 & 1.58 & 0.051 & RE19\\
STIS & 2010-02-04 & 0.37 & 2.31e-14 & 2.33 & 0.019 & RE19\\
XSH & 2010-04-07 & 0.75 & 3.51e-14 & 1.70 & 0.026 & V19\\
XSH & 2010-05-03 & 0.59 & 3.85e-14 & 2.07 & 0.029 & M14\\
STIS & 2010-05-28 & 0.85 & 4.68e-14 & 1.48 & 0.031 & RE19 \\
STIS & 2015-04-18 & 0.89 & 5.31e-14 & 1.66 & 0.041 & RE19 \\
HIRES & 2008-05-23 &  0.43 & 3.01e-14 & 2.37 & 0.024 & F18\\
SNIFS & 2014-11-27 & 1.22 & 6.10e-14 & 1.33 & 0.051 & G18\\  %331
SNIFS & 2014-11-29 & 1.94 & 1.21e-13 & 1.52 & 0.081 & G18\\  %333
SNIFS & 2014-12-08 & 0.72 & 4.47e-14 & 2.47 & 0.042& G18\\  %342
SNIFS & 2014-12-10 & 1.56 & 6.43e-14 & 1.37 & 0.050& G18\\  %344
SNIFS & 2014-12-13 & 1.22 & 6.79e-14 & 1.72 & 0.049 & G18\\  %347
XSH & 2019-07-06 & 1.70 & 8.46e-14 & 1.43 & 0.061 & -- \\
XSH & 2019-07-07 & 1.86 & 1.21e-13 & 1.47 & 0.105 & -- \\
XSH & 2021-04-02 & 1.27 & 7.71e-14 & 1.60 & 0.055 & M21\\
XSH & 2021-04-06 & 1.70 & 1.15e-13 & 1.45 & 0.092 & M21\\
XSH & 2021-04-08 & 0.22 & 2.47e-14 & 2.60 & 0.020 & M21\\
\hline
\multicolumn{7}{l}{$^a$Veiling at 5050 \AA}\\
\multicolumn{7}{l}{$^b$Accretion continuum flux at 5050 \AA.}\\ % in $10^{-14}$ \erga.}\\
\multicolumn{7}{l}{$^c$Accretion luminosity in L$_\odot$}\\
\multicolumn{7}{l}{References:  H04: \citet{herczeg04}}\\
\multicolumn{7}{l}{~~~~HH08: \citet{Herczeg2008}}\\
\multicolumn{7}{l}{~~~~H09: \citet{Herczeg2009}}\\
\multicolumn{7}{l}{~~~~HH14:  \citet{herczeg14} and in prep.}\\
\multicolumn{7}{l}{~~~~RE19:  \citet{robinson19}}\\
\multicolumn{7}{l}{~~~~F18:  \citet{fang18}}\\
\multicolumn{7}{l}{~~~~V19:  \citet{venuti19}}\\
\multicolumn{7}{l}{~~~~M14:  \citet{manara14}}\\
\multicolumn{7}{l}{~~~~G18:  \citet{guo18}}\\
\multicolumn{7}{l}{~~~~M21:  \citet{manara21}}\\
\end{tabular}
\label{tab:lowres}
\end{table}

\begin{table}[!b]
\caption{Comparison of accretion luminosities}
\begin{center}
\begin{tabular}{ccccccc} \hline\hline
 & & Here &  Lit. & \\
Instr. & Date & $L_{acc}$ &  $L_{acc}$ & Reference \\
\hline
STIS & 2010-01-28 & 0.051 & 0.071 & RE19\\
STIS & 2010-02-04 & 0.019 & 0.030 & RE19\\
STIS & 2010-05-28 & 0.031 & 0.048 & RE19\\
STIS & 2015-04-18 & 0.043 & 0.057 & RE19\\
XSH & 2010-05-03 & 0.029  & 0.030 & M14\\ 
XSH & 2010-04-07 & 0.026 & 0.027 & V19\\
\hline
\multicolumn{5}{l}{RE19: \citet{robinson19}}\\
\multicolumn{5}{l}{M14: \citet{manara14}}\\
\multicolumn{5}{l}{V19: \citet{venuti19}, 60s spectrum}\\
\end{tabular}
\end{center}
\label{tab:mdotcomp}
\end{table}

\begin{figure*}[!t]
\centering
\includegraphics[trim={2.4cm 13cm 2.2cm 1.5cm},width=0.48\textwidth]{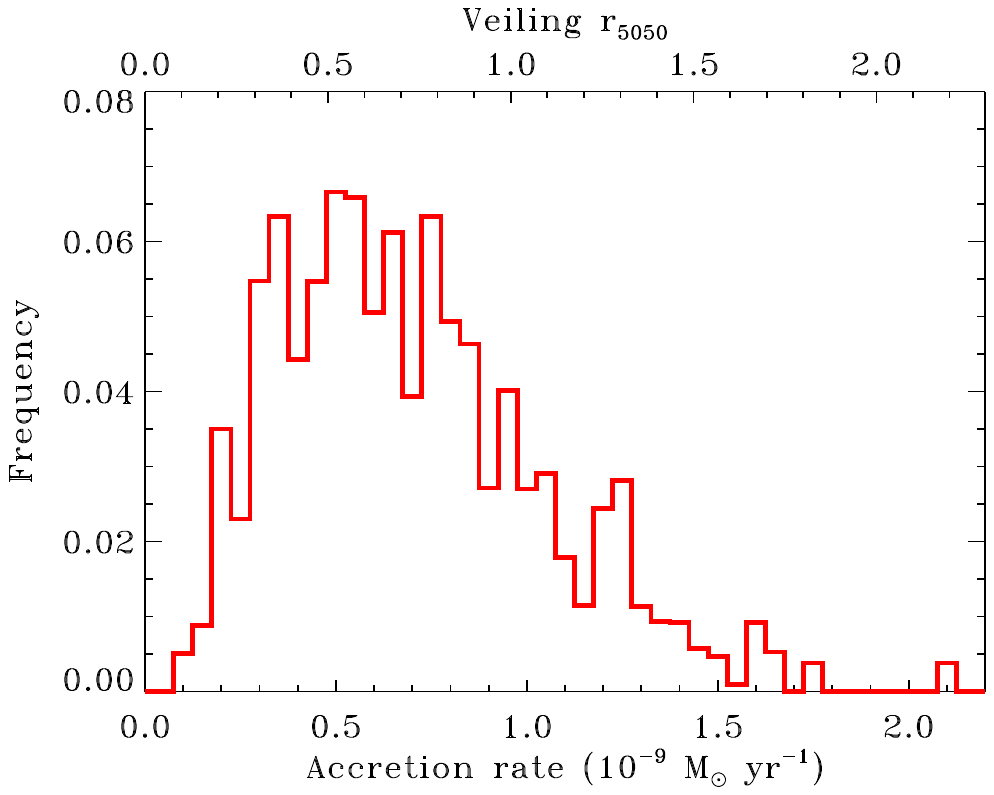}
\includegraphics[trim={2.4cm 13cm 2.2cm 1.5cm},width=0.48\textwidth]{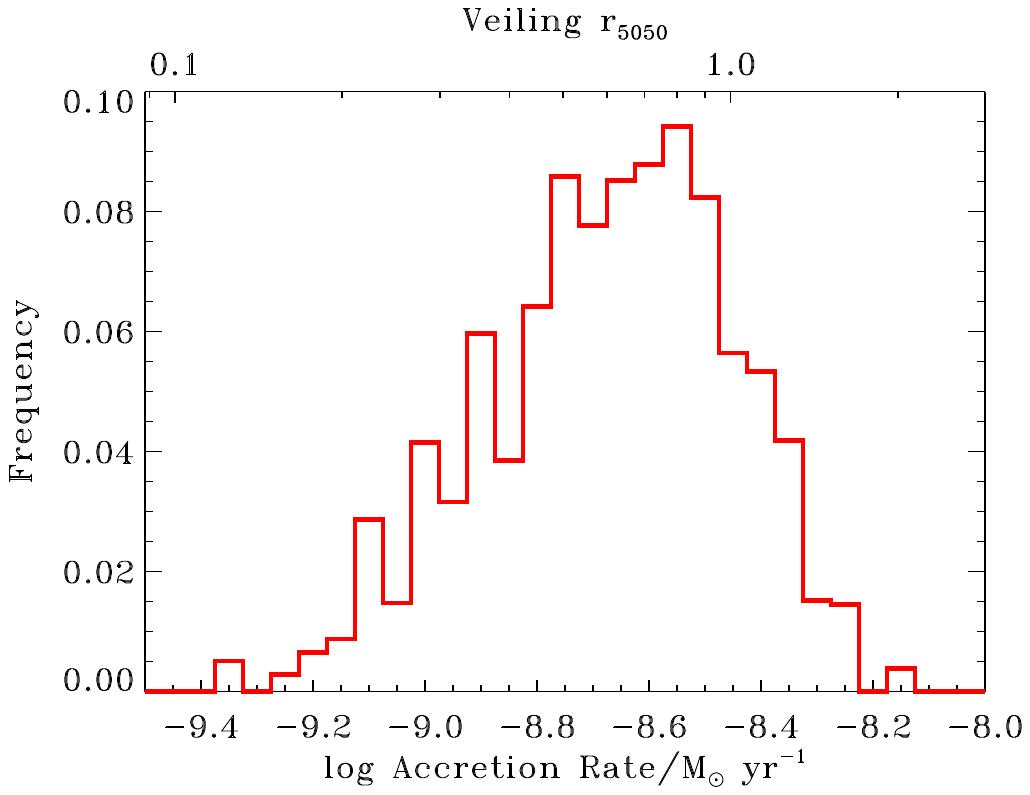}
\caption{Histograms of accretion rates, plotted as a distribution in linear (left) and logarithmic (right) values, with corresponding veiling on the top axis.   The average $\log \dot{M_{\rm acc}}$ is $10^{-8.65}$~M$_\odot/yr$ with a standard deviation of $0.22$ dex.  The differences in veiling are much larger than the precision and cross-instrument uncertainties of $\sim 0.02$. }
\label{fig:histogram}
\end{figure*}

\subsection{Final Calculation of Continuum-based Accretion Rates}

Figure~\ref{fig:histogram} shows a histogram of accretion rates, as measured from the continuum luminosity.  In our time series from 1998--2022, the average veiling is 0.73 with a standard deviation of 0.36, the average
accretion luminosity is 0.045 $L_\odot$ (standard deviation of 0.022 $L_\odot$),
and average accretion rate is $2.51\times10^{-9}$ M$_\odot$ yr$^{-1}$ (standard deviation $1.25\times10^{-9}$); the average $\log \dot{M_{\rm acc}}/M_\odot {\rm yr}^{-1}$ is $-8.65$ (standard deviation 0.22).  The full range of veiling spans from 0.13 to 2.13, corresponding to accretion rates from $0.46$--$7.4\times10^{-9}$~$M_\odot$~yr$^{-1}$.  The veiling never goes below zero or even close to the uncertainty in the zero point, so excess continuum emission due to accretion is detected in every spectrum.

These averages are calculated after reducing the contributions of observations that were obtained contemporaneously.  Veiling measurements are randomly selected into a sample.  When one point is selected, all other points obtained within 3 hrs are excluded.  The final histogram and averages are obtained by averaging 10,000 different selections.  An average without any weighting would lead to slightly higher measurements because many Magellan/MIKE spectra were obtained on a night when the veiling was high.  The average of the logarithm of accretion rate leads to a slightly smaller accretion rate than the linear average.

The variability of the log of the accretion rate is $\sim 0.22$~dex, as found here from the distribution and also in \S7.2 from a structure function analysis.  This variability is consistent with the variability found previously in most single-object analyses and surveys \citep[e.g.][]{biazzo12,Costigan2014,zsidi22,fiorellino22}, though EX Lup-type objects have bursts that are much larger than found in generic samples \citep[see review by][]{fischer22}.  The factor of $90$ change in the accretion rate of XX Cha \citep{claes22} is also larger than the full range of accretion rates measured here for TW Hya, so some accreting young stars seem to have accretion rates that are more variable than that of TW Hya.

\section{Line emission and accretion rate variability}
\label{sec_lines}

The luminosity of emission lines in CTTS spectra is often used to measure accretion through correlations with accretion luminosity.  These correlations were developed with single-epoch spectra from a large number of stars and 
are sometimes applied to spectral monitoring of spectroscopic features of individual stars to infer variability.  However, the response of lines to changes in accretion rate for an  individual star does not necessarily follow the same relationship as found for the global correlations obtained in large samples of stars.  

\begin{figure*}[!t]
\centering
\includegraphics[trim={2.5cm 13cm 2.5cm 3cm},width=0.33\textwidth]{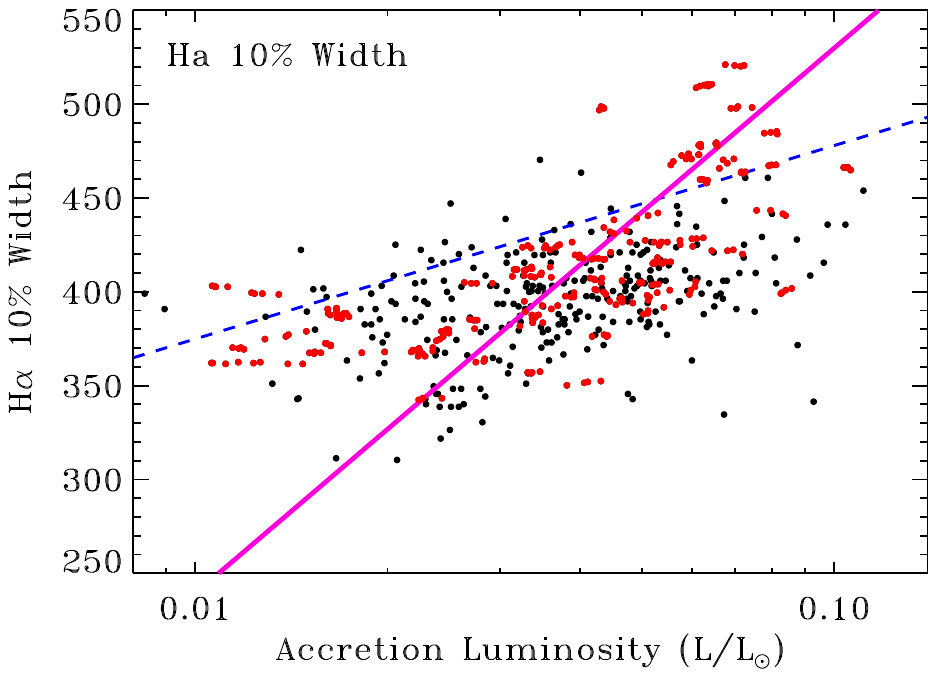}
\includegraphics[trim={2.5cm 13cm 2.5cm 3cm},width=0.33\textwidth]{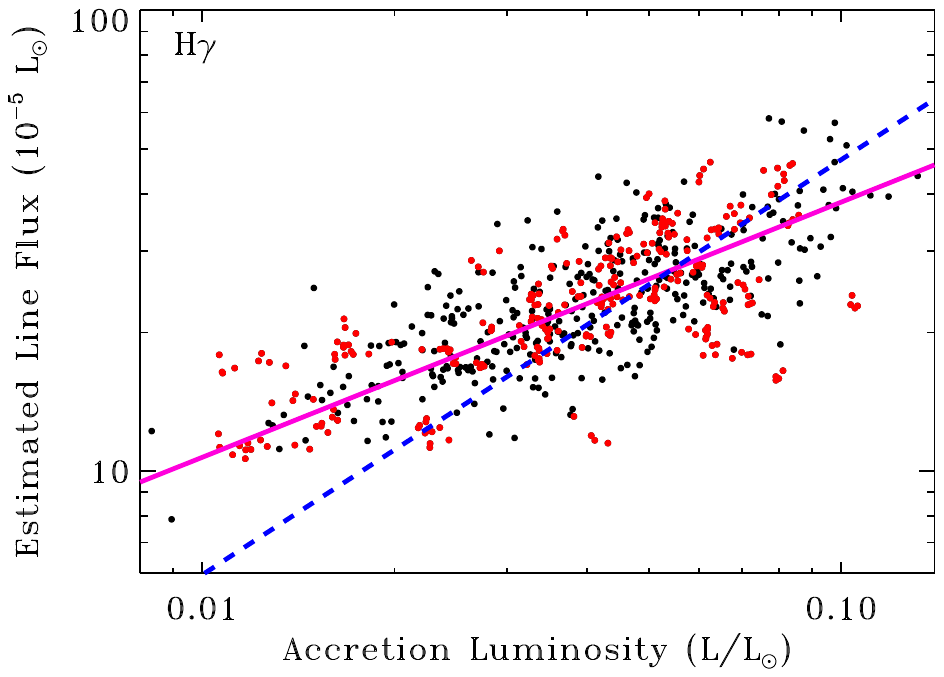}
\includegraphics[trim={2.5cm 13cm 2.5cm 3cm},width=0.33\textwidth]{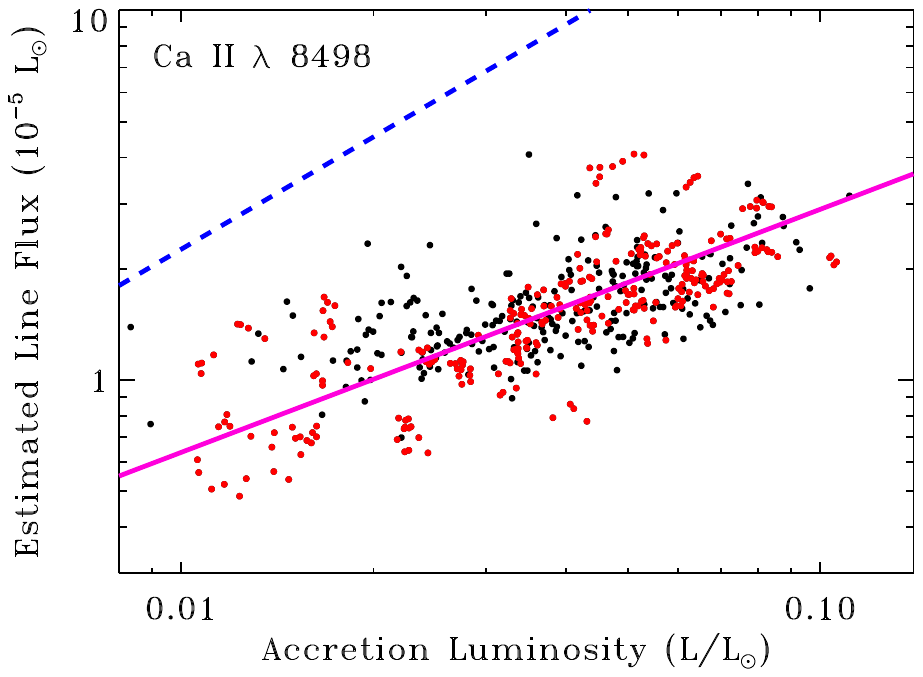}
\includegraphics[trim={2.5cm 13cm 2.5cm 3cm},width=0.33\textwidth]{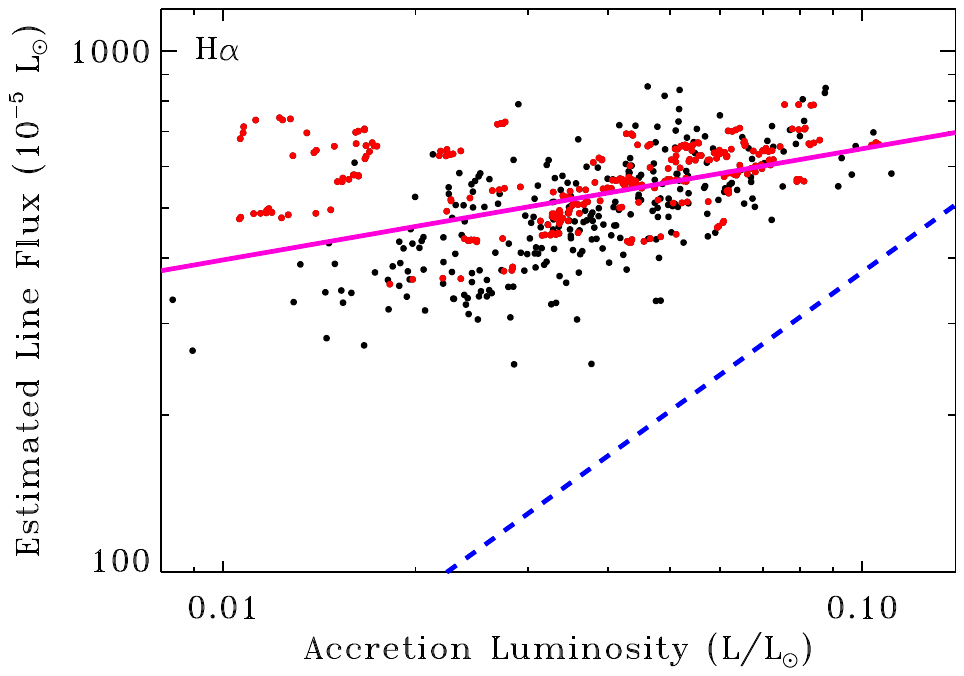}
\includegraphics[trim={2.5cm 13cm 2.5cm 3cm},width=0.33\textwidth]{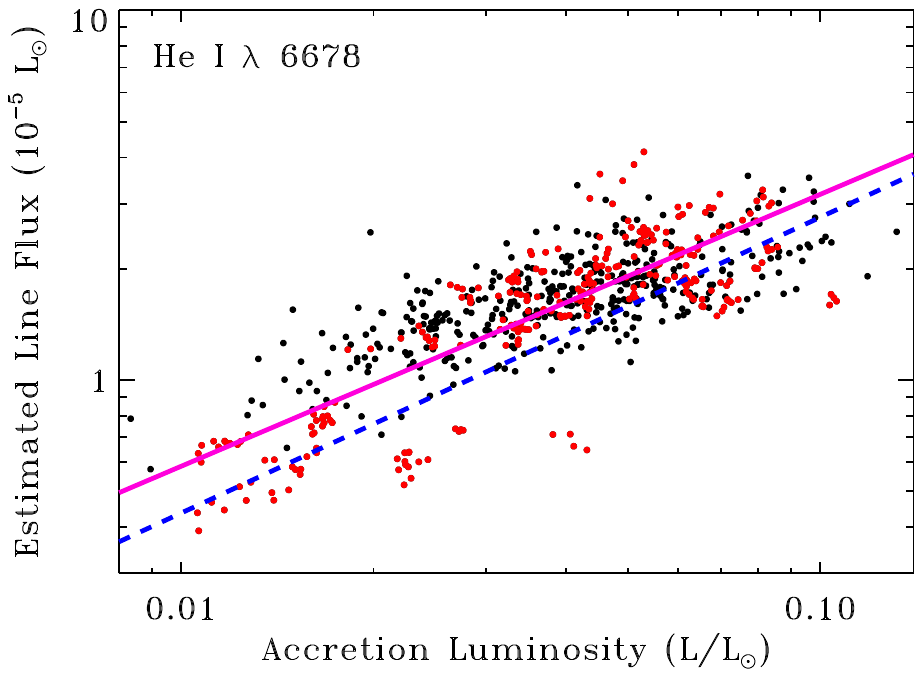}
\includegraphics[trim={2.5cm 13cm 2.5cm 3cm},width=0.33\textwidth]{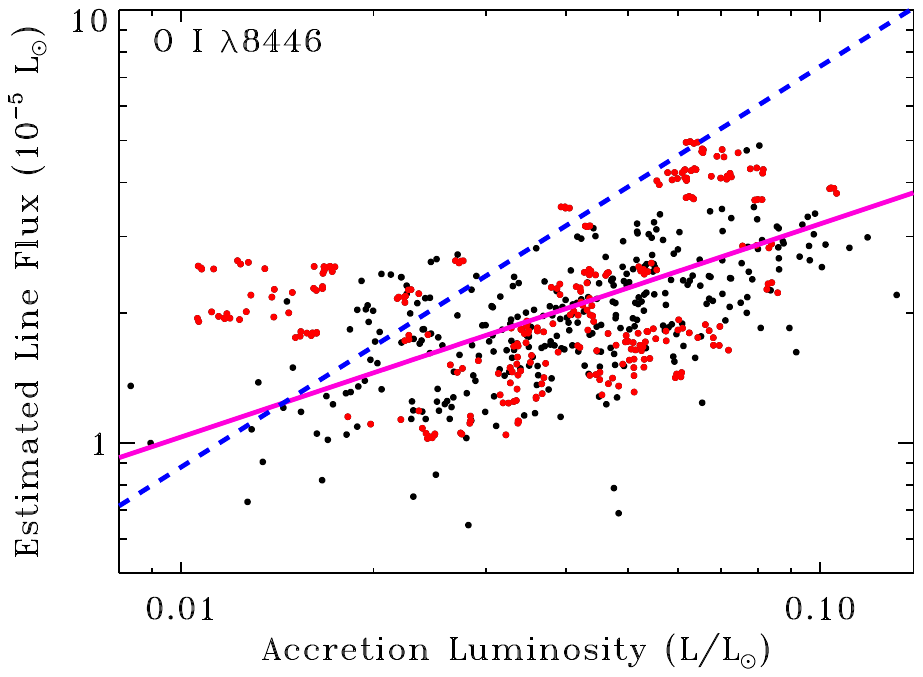}
\includegraphics[trim={2.5cm 13cm 2.5cm 3cm},width=0.33\textwidth]{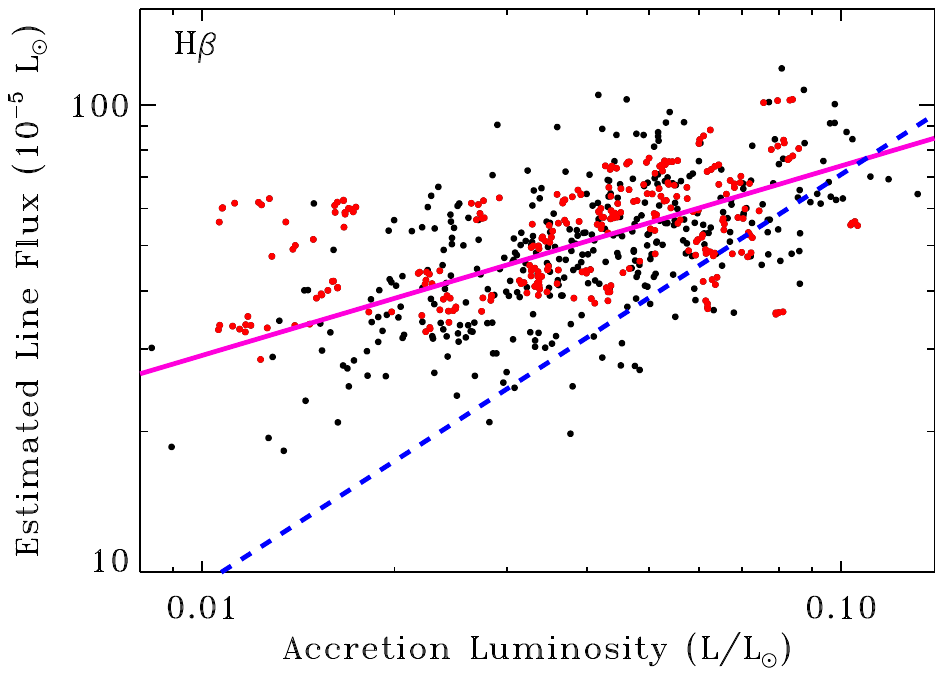}
\includegraphics[trim={2.5cm 13cm 2.5cm 3cm},width=0.33\textwidth]{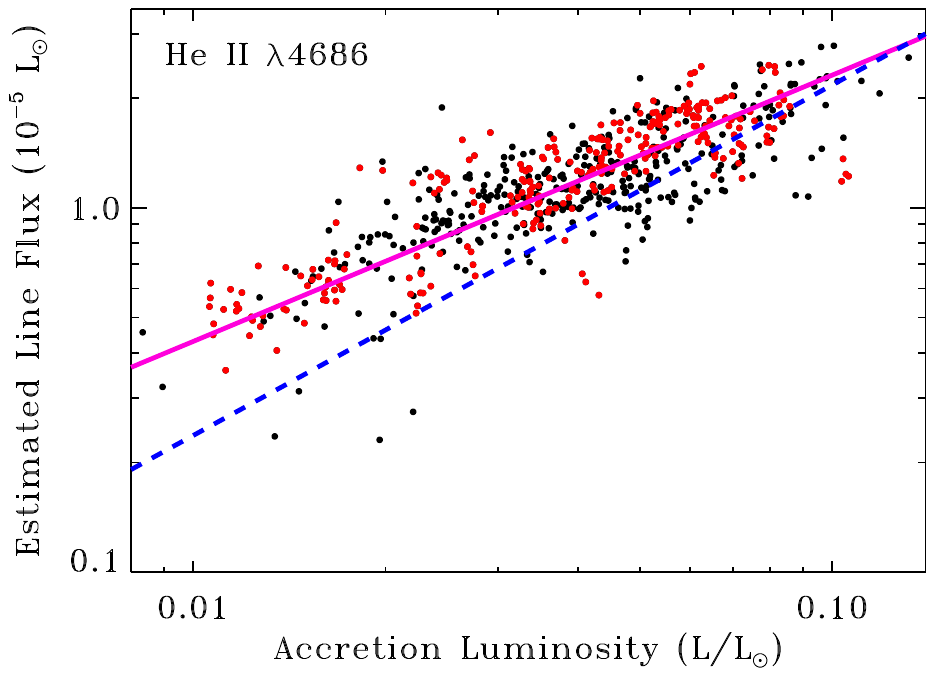}
\includegraphics[trim={2.5cm 13cm 2.5cm 3cm},width=0.33\textwidth]{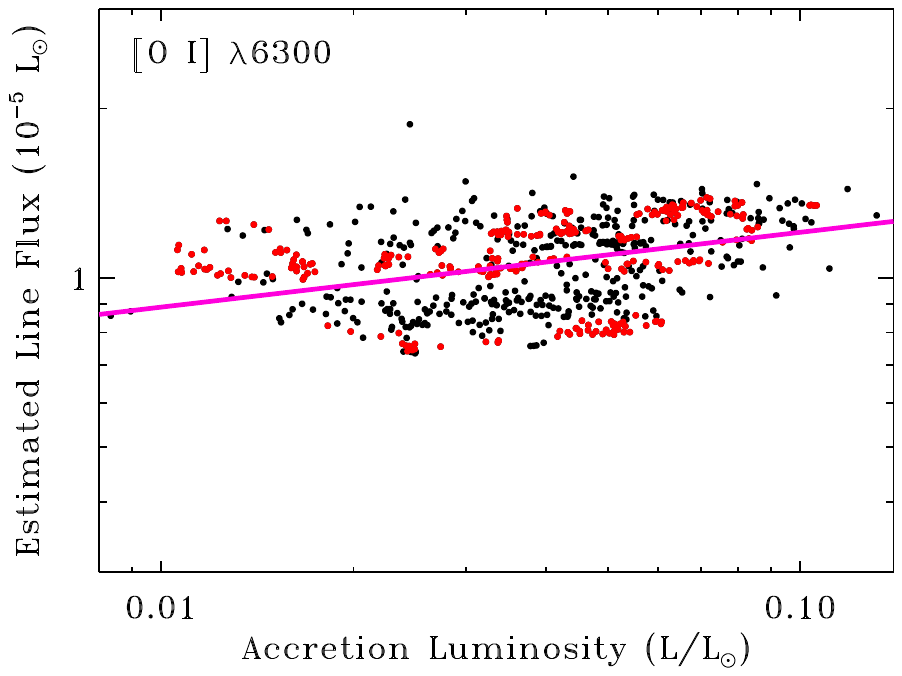}
\caption{Correlations between accretion luminosity and line luminosities (red circles from ESPaDOnS, black circles from other instruments), with a best-fit (solid pink line) and a comparison to the best-fit correlation (dashed blue line) for the relationships from \citet{alcala17}.  For H$\alpha$ 10\% width, the comparison is to the correlation calculated by \citet{Natta2004}, adjusted to accretion luminosity given the parameters for TW Hya.  Some of the lines are poorly correlated with the accretion luminosity.}
\label{fig:linecorrs}
\end{figure*}

In this section, we study the reliability of these correlations as a variability indicator for a single star, TW Hya.  Correlations between line and accretion luminosities are described in Table~\ref{tab:linecorrs} and Figure~\ref{fig:linecorrs}.  
The lines are selected to avoid severe blends.  We also avoid lines located at wavelengths shorter than 4300 \AA, where high veilings lead to uncertain line luminosities.
The \ion{Ca}{2} $\lambda8498$ line is selected as the member of the \ion{Ca}{2} infrared triplet that is least blended with Paschen lines.  Paschen-10 at 8598 \AA\ is selected as the lowest Paschen line in optical spectra that is not affected by telluric absorption and is not blended with any of the \ion{Ca}{2} infrared triplet lines.  The line emission is integrated over the entire profile, despite narrow and broad components that may respond differently to changes in accretion rate.   We restrict this analysis to ESPaDOnS, FEROS, HARPS, UVES (4900--7000 \AA), ESPRESSO, CHIRON, and HIRES spectra, since they have the most reliable measurements and consistent set of reductions.  

As with the veiling measurements, this analysis assumes a constant photospheric flux.  Most of these lines are also be seen in emission in chromospherically active stars.  The equivalent width are measured only after subtracting off a chromospheric template, which provides a natural correction for the chromospheric emission.  The chromospheric line emission is much fainter than line emission from accretion and will have a negligible affect on this analysis.

\begin{table}[!b]
\centering
\caption{Empirical Parameters of Accretion Rate Indicators$^a$}
\begin{tabular}{lcccccc}
\hline
Line & $a$ & $\sigma(a)$ & $b$ & $\sigma(b)$ & $\sigma$ & P$^b$ \\
\hline
H$\alpha$ $\lambda6563$ &  9.3 & 1.0 & 4.7 & 0.4 & 0.41 & 0.41\\
H$\alpha$ 10\% width & -2.82 & 0.05 & 0.0034 & 0.0001 & 0.18 & 0.60\\
H$\beta$ $\lambda4860$  & 6.72 & 0.38 & 2.47 & 0.11 & 0.29 & 0.53\\
H$\gamma$ $\lambda 4340$ & 5.18 & 0.24 & 1.81 & 0.07 & 0.19 & 0.72\\
Paschen-10 $\lambda8598$ & 2.91 & 0.12 & 0.85 & 0.02 & 0.20 & 0.58\\
\ion{He}{1} $\lambda4471$ & 4.39 & 0.17 & 1.23 & 0.04 & 0.17 & 0.75\\
\ion{He}{1} $\lambda5876$ & 4.45 & 0.17 & 1.30 & 0.04 & 0.18 & 0.69\\
\ion{He}{1} $\lambda6678$ & 5.11 & 0.20 & 1.36 & 0.04 & 0.16 & 0.77\\
%%%\ion{He}{1} $\lambda7065$ & \\
\ion{He}{2} $\lambda4686$ & 5.35 & 0.21 & 1.37 & 0.04 & 0.14 & 0.83\\
\ion{O}{1} $\lambda6300$  & 36   & 5    & 7.5 & 1.0 & 0.54 & 0.35 \\
\ion{O}{1} $\lambda8446$  & 8.12 & 0.40 & 2.03 & 0.09 & 0.30 & 0.46\\
\ion{Ca}{2} $\lambda8498$ & 5.90 & 0.27 & 1.52 & 0.06 & 0.18 & 0.71\\
%%%\ion{Na}{1} $\lambda 5896$ & \\
\hline
\multicolumn{7}{l}{$^a$$\log L_{\rm acc}=a+b\times \log L_{\rm line}$ (or $\log L_{\rm acc}=a+b\times 10\%$ width).}\\
\multicolumn{5}{l}{$^b$Pearson correlation coefficient}\\
\end{tabular}
\label{tab:linecorrs}
\end{table}

\subsection{Line Luminosity Measurements}

Line luminosities are estimated by first measuring veiling-corrected equivalent widths and subsequently converting them to luminosities from flux-calibrated photospheric templates. The equivalent width of each line is measured by integrating emission in the line profile after subtracting off a high resolution spectrum of the photospheric template TWA 25.  The measured equivalent width is then corrected for veiling by multiplying by $1+r_\lambda$ and converted to a flux using the scaled photospheric template.  This correction is necessary here but not in other studies \citep[e.g.][]{alcala17}, because most of the high-resolution spectra presented here are not flux-calibrated.
The assumption in our calculation is that the photospheric emission is constant.

When subtracting the photosphere, the spectra of TW Hya are first rotationally broadened to match the $v\sin~i$ of TWA 25.  Then either the TWA 25 or TW Hya spectrum (depending on the instrument resolution) is broadened with a Gaussian profile to account for the minor differences in resolution between, for example, HARPS and ESPaDOnS, so that the line widths match.  The TWA 25 spectra are then shifted to the velocity of TW Hya.  
The stellar photosphere is then scaled to each line location based on the veiling measurement for that spectrum at 5000-5100 \AA.  The veiling at the respective wavelength is calculated based on the wavelength dependence developed from the ESPaDOnS spectra in \S 3.

For most lines, the equivalent width is calculated by integrating emission from $\pm 500$ \kms, wide enough to include emission in the line wings.   The continuum spans 5--10 \AA\ intervals around each line, selected to avoid strong features. For [\ion{O}{1}] $\lambda6300$, the equivalent width is measured from a best-fit Gaussian profile to the line.
 
 For H$\alpha$, the integration region\footnote{The [\ion{N}{2}] $\lambda 6548$ and $\lambda6583$ lines are not detected and do not contribute to these measurements.} covers $\pm 1950$ \kms\ from line center because of the extreme line width (see \S 6.4.1), while the continuum region is measured from $\sim 3000-4000$ \kms.  The 10\% width for the H$\alpha$ line is measured by smoothing the line profile by $\sim 15$ \kms\ and calculating the location where the continuum-subtracted flux exceeds 10\% of the peak value.   For analysis of H$\alpha$, we exclude all UVES and CHIRON spectra plus 13 FEROS and HARPS spectra with an H$\alpha$ peak that is at or near saturation.

The [\ion{O}{1}] $\lambda6300$ line may be contaminated by telluric emission.  The analysis excludes CHIRON, which has no sky subtraction and is at a spectral resolution low enough that the [\ion{O}{1}] sky emission blends with the source emission.

\begin{figure*}[!t]
\centering
\includegraphics[trim=30mm 22mm 0mm 175mm,width=0.98\textwidth]{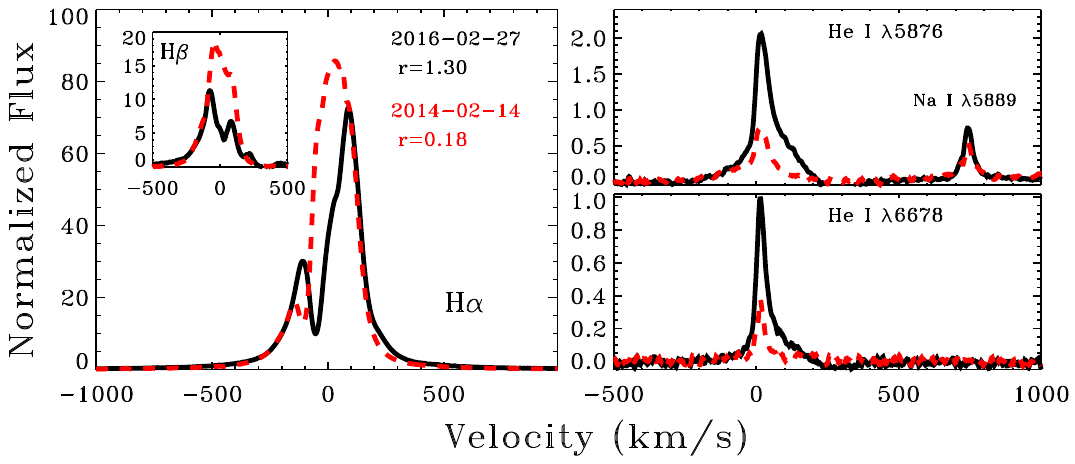}
\caption{H$\alpha$, H$\beta$, and two \ion{He}{1} lines from an epoch with strong accretion (solid black line) and an epoch with weak accretion (dashed red line).  The spectra shown here have had the photosphere subtracted and are then normalized to the relevant photospheric continuum to place the lines on the same flux scale.  The He lines are brighter but the H lines are fainter during the epoch with stronger accretion, illustrating by example that accretion is more tightly correlated with He emission than with H emission.  The H lines are formed in an optically thick medium and have emission that may be absorbed by the wind or accretion flow.}
\label{fig:lines_hilo}
\end{figure*}

\subsection{Relationships between line and accretion luminosities}

The correlations between veiling and line luminosities and equivalent widths are presented in Table~\ref{tab:linecorrs} and Figure \ref{fig:linecorrs}, with linear fits to $\log L_{\rm acc}=a \times b \log L_{\rm line}$.  The quality of these relationships are evaluated from the standard deviation of the difference between measured $\log L_{\rm acc}$ and that estimated from the best-fit line and also from the Pearson correlation coefficient.  The standard deviation of all measured $\log L_{\rm acc}$ is $\sim 0.22$ dex, so any useful relationship would need to predict the $\log L_{\rm acc}$ with a standard deviation significantly lower than $0.22$ dex.  As a general rule, the Pearson correlation coefficient is higher than 0.7 for strong correlations and less than 0.4 for poor or no correlation.

Most lines provide some limited predictive power.  The He lines are well correlated with accretion luminosity, with best-fit relationships that have significant power above random noise.  The H$\gamma$ and Paschen-10 line luminosities are also well correlated with accretion and provide some limited predictive power.    Presumably H lines from even higher energy levels also scale well with accretion.  H$\beta$ is well correlated with accretion luminosity, though the scatter is higher than that for other H lines and the correlation has a different slope than that measured in the sample of accretors by \citet{alcala17}.
The luminosity of H$\alpha$ is only poorly correlated with accretion luminosity and offers no predictive power.  The 10\% width of H$\alpha$ is well correlated with accretion luminosity but with a relationship that provides only modest power.  Neither the \ion{O}{1} $\lambda8446$ line, likely produced by the heated chromosphere, nor the [\ion{O}{1}] $\lambda6300$ line (see \S 6.3.3) are correlated with accretion variability.

The \ion{Ca}{2} line emission correlates with accretion luminosity but is much weaker than expected from the \citet{alcala17} relationship, which may be related to disk structure \citep{micolta23}.  The accretion flow of TW Hya is deficient in elements such as Fe and O (e.g., \citealt{Kastner2002}; see also speculation about Si depletion by \citealt{Herczeg2002}) and may likewise be deficient in Ca.  As described in models of \citet{kama16} and \citet{booth18}, volatile and refractory species may be preferentially in ices and grains and not in accretion flows.  However, models of the accretion flow and shock are needed to accurately measure abundances to determine whether the weak \ion{Ca}{2} lines are best explained by an underabundance or by the physical conditions of the emission region.

To illustrate the utility of He lines and futility of H lines \citep[see also phrasing from][]{fischer22} as accretion rate indicators, Figure~\ref{fig:lines_hilo} compares H and He lines from a strong accretion spectrum from 2016-02-27 with a weak accretion spectrum from 2014-02-14.  The He lines are all stronger during the high accretion rate epoch, as expected, while the H$\alpha$ lines are actually stronger during the epoch of low accretion rate.  The difference in the H$\alpha$ line flux is explained by high line opacity and absorption in the winds and accretion flows.  However any empirical correction in the line profile by, for instance, scaling the weak accretion spectrum to fill in missing flux near line center, would only lead to the two epochs having similar line fluxes.

For TW Hya, the fractional change in line luminosity is smaller than the fractional change in the accretion luminosity.  The correlations from \citet{alcala17} have slopes of $\sim 1$, consistent with the idea that an increase in accretion rate corresponds to an equal increase in line luminosity.  The \citet{alcala17} correlations were developed from X-Shooter surveys of accreting stars in the Lupus star-forming regions across a wide range of stellar mass.  While those correlations are robust across different objects, applying those correlations to the time-series observations of TW Hya would overestimate the level of variability.

Some scatter may be introduced in the correlations because the response in some lines may be delayed from the continuum \citep{Dupree2012}, as well as by blueshifted absorption in the wind and redshifted absorption in the accretion column.  The H emission line emissitivity extends across the $\sim 3.5$ R$_*$ magnetospheric cavity \citep{gravity_garcialopez20}, so stronger emission in H lines may precede stronger continuum emission by the freefall time, $\sim 0.3$ days.  The H lines may also vary less because the accretion flow and streams are roughly constant, while the accretion shock itself is a near-instantaneous measurement.  With multiple accretion streams, some components and indicators will lag others \citep{espaillat21,robinson22}

 Many line profiles have a broad component from the accretion flow and shock and a narrow component from the heated photosphere \citep[e.g.][]{yang05,Donati2014}.  Variability in line profiles is not analyzed in detail in this paper, despite potential trends of stronger emission in broad profiles during periods of high accretion.  
 However, in \S 6.4 we present some limited results on line profiles.

\subsection{The total luminosity of emission lines}

The accretion rates measured here and in most other studies are calculated from the continuum emission.  Even when emission lines are used, they serve as a proxy for the continuum emission.  The energy that escapes in lines is usually excluded from these calculations, despite significant potential contributions from the lines \citep[see discussion in][]{Alcala2014}.  For TW Hya, this exclusion leads to an underestimate of accretion rates of $\sim 70\%$, based on the following calculations.

At optical wavelengths, the H Balmer series dominates the emission line luminosity.  Because the H$\alpha$ line luminosity is poorly correlated with the accretion (continuum) luminosity,  the ratio the H$\alpha$ luminosity to the accretion luminosity is $\sim 0.5$ at the lowest accretion rates to $\sim 0.05$ at the highest accretion rates.  The average H$\alpha$ luminosity is $0.15$ times the accretion luminosity.  Other optical and near-IR lines, including the Balmer series and \ion{Ca}{2} H \& K lines add another $\sim 0.05$, so this leads to an extra 20\% in accretion.  

The ultraviolet adds even more energy (see Tabulation by \citealt{herczeg04}).  From 1230--3000 \AA, the emission line luminosity is $\sim 8\%$ of the accretion luminosity.  However, the Ly$\alpha$ line dominates the UV flux.  In five observations, the detected Ly$\alpha$ emission is 10\% of the continuum accretion luminosity at high accretion rates and 30\% at low accretion rates.  Moreover, about half of the Ly$\alpha$ flux is unobserved, obscured by circumstellar and interstellar \ion{H}{1}.  Corrections lead to total Ly$\alpha$ fluxes that are roughly two times higher than the observed flux \citep[e.g.][]{herczeg04,schindhelm12,arulananthan23}.  

The X-ray luminosity, as measured by \citep{Kastner2002} and dominated by accretion \citep[e.g.][]{Brickhouse2010,argiroffi17}, adds only $\sim 1\%$ to the total accretion luminosity.

\subsection{Behavior of specific emission lines}

\begin{figure}[!t]
\centering
\includegraphics[trim={2.5cm 13.2cm 2.5cm 3cm},width=0.48\textwidth]{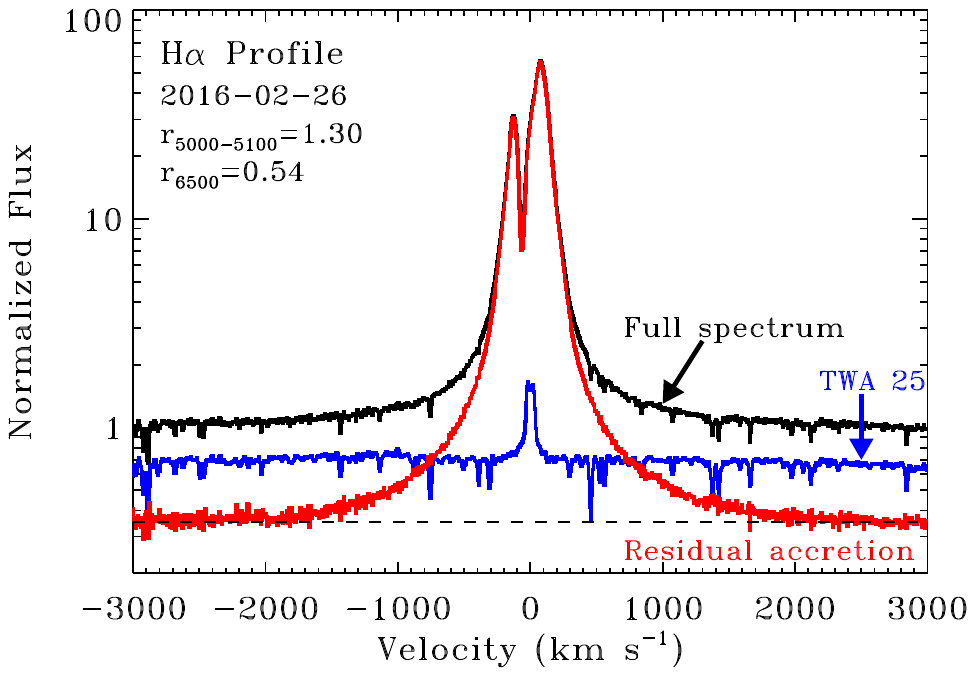}
\caption{The H$\alpha$ profile shows strong Lorentzian wings with emission that extends to $\sim 2000$ \kms\ on both sides of the line profile.  The spectrum shown here (black), from an epoch with strong accretion, is normalized by the photospheric continuum level.  The photospheric template TWA 25 (blue spectrum) is scaled to the appropriate level based on the veiling and is then subtracted from the full spectrum to calculate the accretion spectrum (red).}
\label{fig:halpha_profile}
\end{figure}

\subsubsection{Extremely broad wings on H lines}

Some emission in H$\alpha$ extends to $\pm 2000$ \kms\ (see example in Figure~\ref{fig:halpha_profile}), far beyond the full width at 10\% of the peak flux of $\sim 400$ \kms\ ($200$ \kms\ on both sides of line center).  The weak line wings are consistent with a Voigt profile with a damping parameter of $\sim 40$ \kms.  The total flux outside the 10\% width is typically $\sim 17.5\%$ (with a standard deviation of 2.6\%) of the total flux in the H$\alpha$ line.  These broad wings are seen consistently across all instruments and levels of veiling.  

The Ly$\alpha$ emission from TW Hya has weak wings that extend to even larger velocities, as noted by \citet{france14}.
The fluxes on the prominent line wings are well reproduced by models with superimposed Gaussian profiles to represent the stellar and accretion emission and ISM and outflow absorption  \citep[e.g.][]{herczeg04,schindhelm12} or resonant scattering through an outflowing H I shell \citep[see analysis of DM Tau by][]{arulananthan23}, but those models do not capture the extremely high-velocity wings out to 2000 \kms.
The H$\beta$ line wings extend to $\sim 1200$ \kms, while higher H lines do not have such broad wings \citep[see, e.g.,][]{wilson22}.  This pattern is consistent with expectations for pressure broadening in optically thick lines. 

Very broad (several thousand \kms) wings of H$\alpha$ have also been detected in symbiotic stars \citep{vanwinckel93,ivison94,selvelli00,skopal06} and planetary nebula \citep{arrieta03,miranda22}.  The structure of the objects is similar to T Tauri stars because a hot ionized medium and an optically thick neutral medium coexist. \citet{lee00} suggested that the broad H$\alpha$ wings in symbiotic stars follow the Lorentzian profile if the wings are formed by Raman scattering with atomic hydrogen of UV radiation near Lyman$\beta$. Furthermore, \citet{chang18} showed that the broad Raman wing near H$\alpha$ is broader than that near H$\beta$. Thus, the broad wings near H$\alpha$ in TW Hya might originate from Raman scattering with atomic hydrogen in the optically thick proto-planetary disk, although further calculations are required to test this hypothesis.

\begin{figure}[!t]
\centering
\includegraphics[trim={4.cm 3.cm 4.8cm 12.0cm},width=0.48\textwidth]{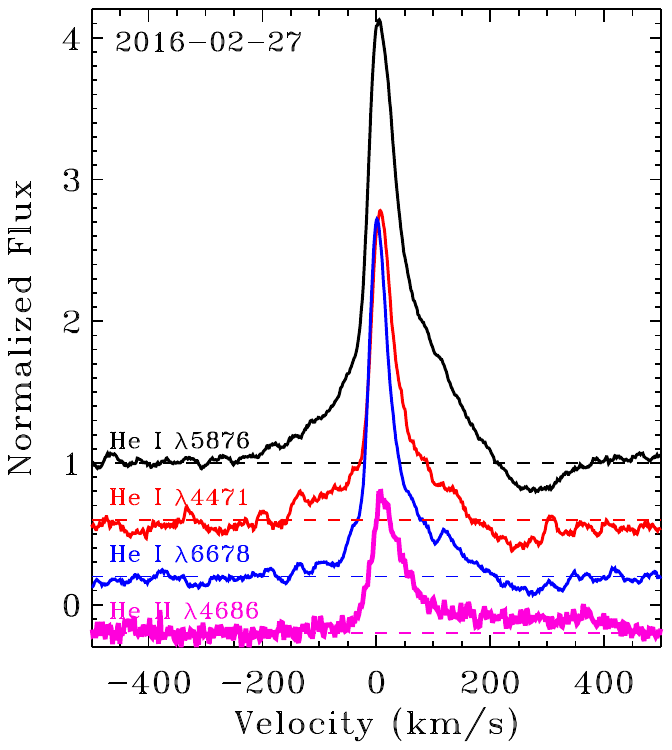}
\caption{\ion{He}{1} absorption, seen here to $\sim 300-400$ \kms\ in three lines on 2016-02-27, an epoch with strong accretion.  The \ion{He}{2} $\lambda4686$ line (bottom, pink) has emission that extends out to 450 \kms.}
\label{fig:he1abs}
\end{figure}

\subsubsection{Inverse P Cygni absorption in \ion{He}{1}}

He lines provide a powerful probe of gas dynamics \citep[e.g.][]{beristain01}.  
Figure~\ref{fig:he1abs} shows that the \ion{He}{1} $\lambda5876$ line exhibits inverse P Cygni profiles during some epochs of strong accretion, with velocities on the red wing that extend to $\sim 350$ \kms.  This detection is challenging because the redshifted absorption blends with the broad wings of photospheric \ion{Na}{1} D absorption.  However, this detection is supported by the detection in other \ion{He}{1} lines of redshifted self-absorption at $\sim 300$ \kms.    For classical T Tauri stars, self-absorption in the optical \ion{He}{1} lines had only been previously identified in a couple of objects \citep{beristain01}.

For comparison, \ion{He}{2} $\lambda4686$ shows emission on the red wing out to $\sim 450$ \kms, similar to the maximum velocities seen in \ion{C}{4} and other high temperature lines \citep[e.g.][]{Herczeg2002,Ardila2013,dupree14}.  This fast velocity must mean that most of the gas is crashing onto the star along our line of sight to the star.  In an accretion geometry with tongues at equatorial or mid-latitudes, the gas would flow more along the plane of the sky, so we would not detect gas at such high velocities.  

Most past work on He lines focused on the \ion{He}{1} $\lambda10830$, which has a metastable lower level that leads to P Cygni (blueshifted) and inverse P Cygni (redshifted) absorption components \citep[e.g.][]{dupree05,edwards06,Fischer2008,erkal22}.  
The lower level of the \ion{He}{1} $\lambda5876$ line is the upper level of the \ion{He}{1} $\lambda10830$, so the absorption in \ion{He}{1} $\lambda10830$ could help to populate the level and lead to the absorption in the $\lambda5876$ line.  A comprehensive analysis of He lines could help to evaluate the excitation and ionization of the accretion flow.

\begin{figure}[!t]
\centering
\includegraphics[trim={2.5cm 13.2cm 2.5cm 3cm},width=0.48\textwidth]{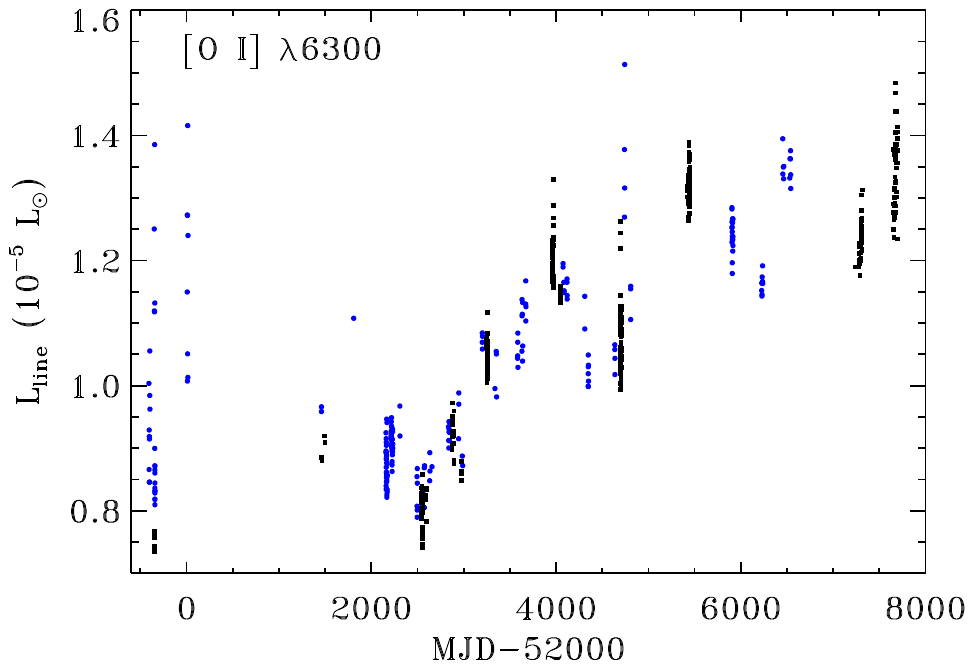}
\caption{The [\ion{O}{1}] $\lambda6300$ line luminosity (blue circles from FEROS, black squares from other instruments) varies with a range of almost a factor of 2 on timescales of months. CHIRON spectra are excluded from this analysis.}
\label{fig:o1lightcurve}
\end{figure}

\subsubsection{Time variability in [\ion{O}{1}] emission}

The [\ion{O}{1}] $\lambda6300$ line profile is narrow and consistent with the low-velocity component of the wind or innermost disk \citep[e.g.][]{fang18,pascucci20}.  Figure~\ref{fig:o1lightcurve} shows that the [\ion{O}{1}] $\lambda6300$ emission varies on timescales of months-years.  The line equivalent width, corrected for veiling, varies from an epoch-averaged minimum of $\sim 0.65$ \AA\ to a maximum of 1.15 \AA, corresponding to luminosities of $(0.8-1.3) \times10^{-5}$ L$_\odot$.

These temporal changes are likely caused by real changes in the line strength.  \citet{fang23} evaluate that the centroid of the narrow line core is stable to within $\sim 1$ \kms\ but find that emission wings that extend to $\sim 30$ \kms\  on both the red and blue sides of the line are sometimes present and sometimes absent. 

Some modest dispersion during each epoch that may be caused by differences in the  spot coverage on the visible surface of the star, since a change in spot coverage will change the flux of the photosphere at 6300 \AA\ but will not change the flux in the line.   However, the long-term temporal changes are too large to be explained by differences in spot coverage.
The ESPaDOnS spectra include telluric [\ion{O}{1}] emission, which in some cases may blend with the stellar emission and in some epochs may be significant.  The shape of the veiling continuum does not vary significantly enough to explain these changes.

\begin{figure*}[!t]
\centering
\includegraphics[trim=30mm 35mm 0mm 56mm,width=0.97\textwidth]{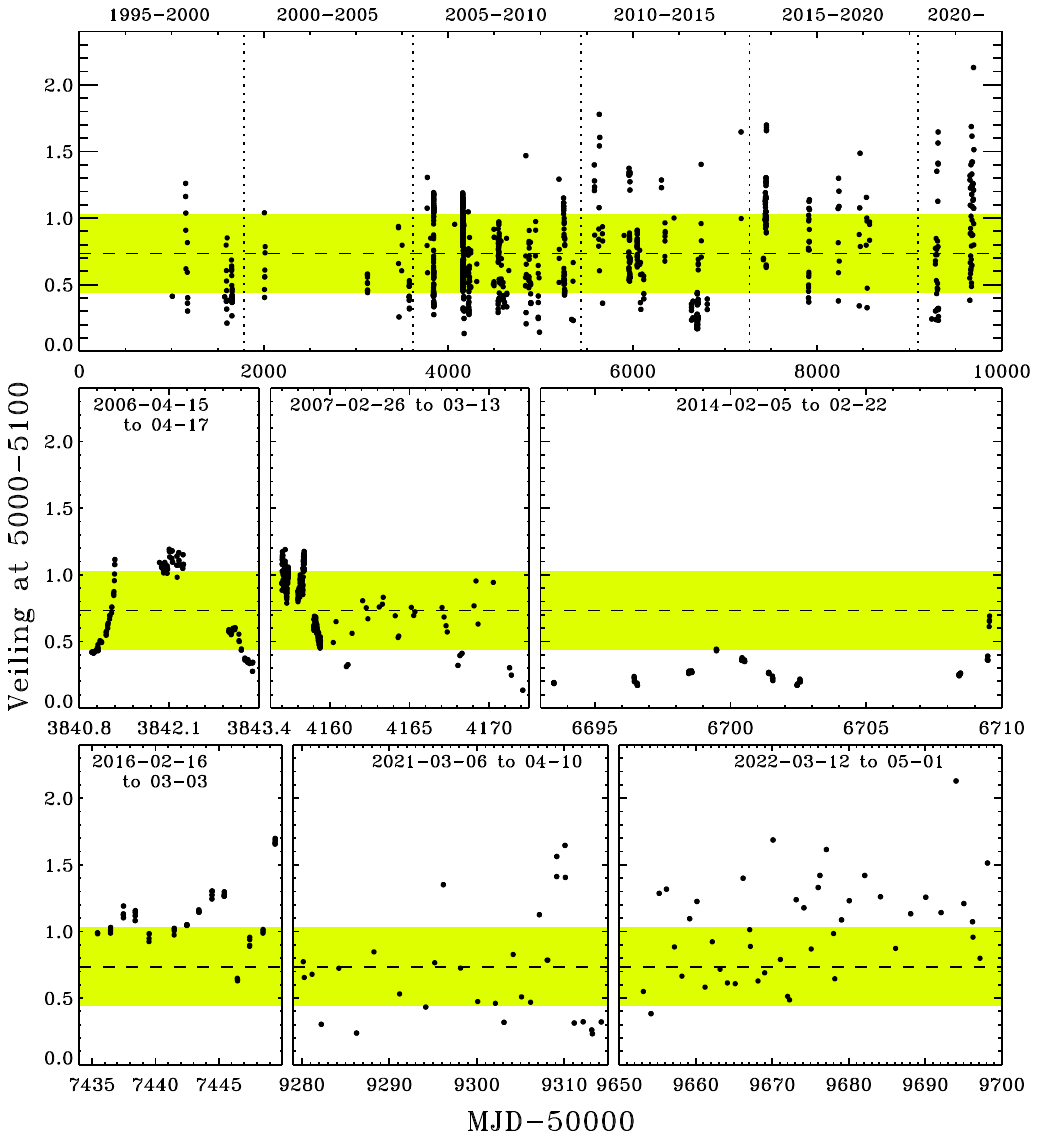}
%\plotone{r5500_eqw}
\caption{The full veiling lightcurve (top), with specific segments highlighted in bottom six panels, with average veiling (horizontal line) and standard deviation (shaded yellow region).  In the six panels, the date range in the x-axis differs in each plot.}
%accounting for differences in resolution and line blending.}
\label{fig:lightcurve}
\end{figure*}

\section{The distribution and variability of accretion onto TW Hya}

\begin{figure}[!t]
\centering
\includegraphics[trim={2.5cm 13.2cm 2.5cm 3cm},width=0.48\textwidth]{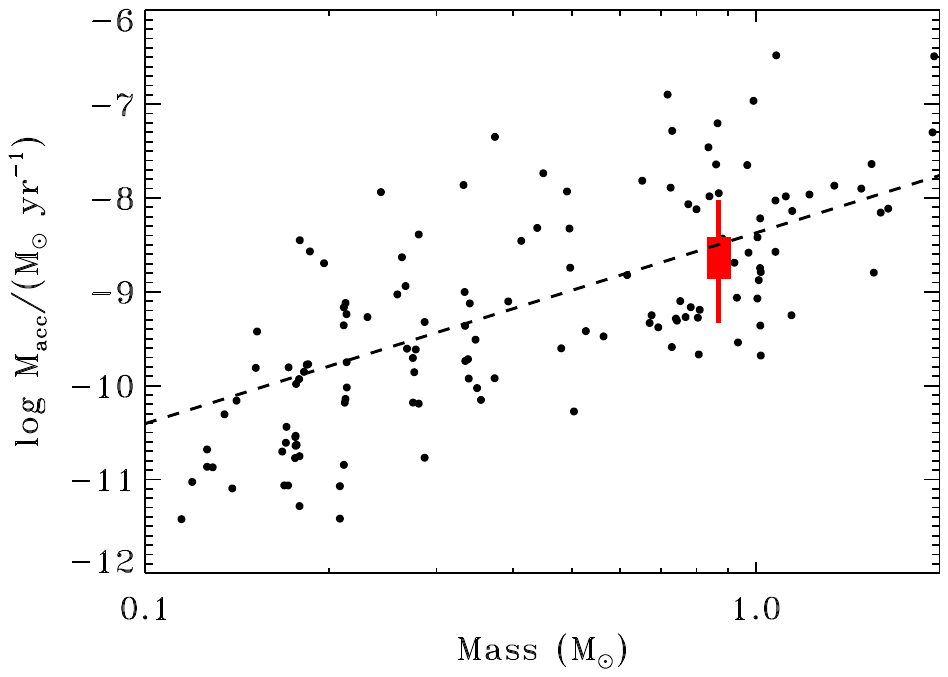}
\caption{Accretion rate of TW Hya (red rectangle), compared with accretion rates from Lupus and Cha I star-forming regions \citep[black circles,][]{alcala17,manara17}.  The size of the thick rectangle for TW Hya shows the $1-\sigma$ scatter of the accretion rates while the thin line shows the full range of accretion rates.  The average accretion rate of TW Hya is only $\sim 0.1$ dex fainter than and consistent with expectations, given the uncertainties, based on a linear fit to the relationship between $\log M_{\rm acc}$ and $\log M_*$.}
\label{fig:mdot_comps}
\end{figure}

\subsection{Average Accretion Rate}% and Interesting Epochs}

The average accretion rate of TW Hya is $2.51\times10^{-9}$~M$_\odot$~yr$^{-1}$, as measured from the continuum emission in \S 5.2.  We adopt this measurement for consistency with previous measurements, however, the exclusion of line emission and the treatment of the accretion flow as a slab rather than a multiple-column flow may cause us to underestimate the accretion rate by a factor of 2--3.

The distribution of $\log$ accretion rate is reasonably well described by a Gaussian profile with an average of $-8.65$, a FWHM of 0.22 dex, and an excess tail at low accretion rates.  In our time series, TW Hya never stops accreting.  The veiling relative to non-accreting templates is always $>0.13$ at 5000--5100, significantly larger than the uncertainty in the zero point.    However, visual inspection of the lightcurve reveals that some epochs have low accretion, including Feb.~2014, and some have high accretion, including Feb.~2016 and Mar.-Apr.~2022.  In addition, the few nights with extensive monitoring show large changes during the night, as originally reported for Magellan/MIKE monitoring \citep{Dupree2012} and seen here also in UVES monitoring.  
Much of the nightly variation seen across the lightcurve may be explained by hours-long bursts, seen in a few high-cadence epochs here as well as in photometric monitoring with MOST \citep[e.g.][]{siwak14,siwak18}.

TW Hya has often been considered a very weak accretor, an interpretation that is not supported by our comparison to unbiased datasets.  Figure~\ref{fig:mdot_comps} compares the accretion rate from TW Hya with accretion rates measured for complete samples of stars in Lupus and Cha~I \citep{manara17,alcala17}, updated with all stellar properties calculated with the Gaia DR3 distance \citep[inverted parallaxes from][]{gaia21} and placed on the mass scale from \citet{somers20} tracks with 50\% spots.  The accretion rate of TW Hya is only 0.15 dex lower than the median accretion rate expected for a $\sim 0.87$ M$_\odot$ star in Lupus or Cha~I, despite the much older age of TW Hya.  

The misconception that TW Hya is a weak accretor has two explanations.  First, the accretion models of \citet{Muzerolle2000} fit to the H$\alpha$ line yielded an accretion rate of $5\times10^{-10}$ M$_\odot$~yr$^{-1}$, 0.75
dex lower than our median accretion rate. 
\citet{Muzerolle2000} measured a veiling of $\sim0.2$ at 7000 \AA, consistent with our veiling measurement\footnote{Measured for lines from $\sim 6200-6500$ \AA; the spectrum does not cover wavelengths below the $6000$} of %0.31 before shifts
0.49 at 5000-5100 \AA\ and yields an accretion rate of $1.69\times10^{-9}$ M$_\odot$ yr$^{-1}$, or 0.17 dex lower than the average accretion rate.  
The remaining difference of $\sim 0.6$ dex between the \cite{Muzerolle2000} and our accretion rates is caused by methodological differences, with our measurements from the accretion luminosity and \citet{Muzerolle2000} from models of the H$\alpha$ line profile.  This difference
is consistent with the offset in the correlation between accretion luminosity and the H$\alpha$ 10\% width for TW Hya, in comparison to the relationship developed by \citet{Natta2004}, as seen in Figure~\ref{fig:linecorrs}.  Second, TW Hya may have also been considered a weak accretor because early accretion rates were measured for biased samples.
 The median accretion rate is $9.6\times10^{-9}$  M$_\odot$ yr$^{-1}$ in \citet{Gullbring1998} and $10.3\times10^{-9}$ M$_\odot$ yr$^{-1}$ in \citet{Valenti1993}, a factor of 3--5 higher than expected for $0.6-1.0$ M$_\odot$ stars from the complete surveys in Lupus and Cha I by \citet{alcala17} and \citet{manara17}.

Our average accretion rate is remarkably consistent with the accretion rate of $1.67\times10^{-9}$~M$_\odot$~yr$^{-1}$ derived from X-ray spectroscopy (\citealt{Brickhouse2012}, increased here by 1.11 to account for the slight change in adopted distance).  These X-ray estimates use hydrodynamic models and account for the H I column density that attenuates the shock, with an interpretation supported by the measurement of redshifted X-ray lines by \citet{argiroffi17}.  The accretion rate measured in the third X-ray integration is at the low end of our distribution but coincides with epochs with low veiling, with the X-ray-based accretion rate a factor of two lower than our measurements.  This agreement is remarkable, since the X-ray method depends only on line ratio diagnostics from He-like \ion{Ne}{9} for electron density, electron temperature, and absorption column density and is independent of 
the X-ray luminosity.
Previous estimates yielded lower accretion rates because the H I column density was underestimated \citep[e.g.][]{stelzer04,guenther07}.

The accretion luminosity is calculated from the continuum alone and excludes emission lines (see \S 6.3).  The accretion luminosity would be $\sim 70$\% higher if emission lines, in particular  Ly$\alpha$ and H$\alpha$, were included as accretion luminosity.  The H line luminosities are relatively steady compared to the fluctuations in the continuum accretion luminosity, which means that this correction is much larger for low accretion rates than for high accretion rates.  Since the H lines are produced over a larger area than the accretion shock \citep{gravity_garcialopez20}, they should provide a more steady source of emission that may not reflect the short-term changes in accretion rate.
The exclusion of lines are even more important towards very low-mass stars and brown dwarfs, since the optical line luminosity to UV continuum luminosity approaches unity \citep{zhou14,Alcala2014}.  The Ly$\alpha$ luminosity for most of those sources is unmeasured and may be higher than H$\alpha$ and other lines in the Balmer series \citep{arulananthan23}.

A second significant uncertainty in our average accretion rate is the bolometric correction, especially in the correction for low-density accretion flows \citep[e.g.][]{Ingleby2013,robinson19,espaillat21}.  This uncertainty affects all epochs of TW Hya and all comparison measurements.  The primary uncertainty in the lower bound of accretion rates is the zero point, where the $\sim 0.02$ absolute veiling uncertainty leads to a $\sim 20\%$ error in the accretion rate.  However, the zero point is a negligible ($<5\%$) uncertainty for epochs of average (or higher) accretion rates.  The width of the accretion rate distribution in Figure~\ref{fig:histogram} is dominated by real variability.

The present-day disk mass within 5 AU is $\sim 7$ M$_J$, as measured from the wings on CO lines uniquely for TW Hya \citep{yoshida22}, and $\sim 50$ M$_J$ for the entire disk \citep{bergin13}.  If the present-day average accretion rate has been constant for 10 Myr, then the total mass accreted onto the star to date would be $\sim 26$ M$_J$; most of the mass was likely accreted when TW Hya was very young.  The initial and even present-day disk mass of TW Hya is near the upper envelope of disk masses around a $\sim 0.87$ M$_\odot$ star, as measured from complete samples of dust emission at mm wavelengths \citep[see comparisons to the compilation by][]{manara22}.  If the accretion onto the star stays steady and the inner disk is not fed additional material from the outer disk, then the inner disk has enough material to survive for another 3.8 Myr.  
The current accretion rate is also a factor of $\sim 10$ higher than the photoevaporation rate of $2-3\times10^{-10}$ M$_\odot$ yr$^{-1}$ \citep{pascucci12}, although the MHD wind may carry significant mass \citep[e.g.][]{wang19}.  These values would all be modified if the line emission increases the accretion rate by 70\%.

\begin{figure*}
\centering
\includegraphics[width=0.9\textwidth]{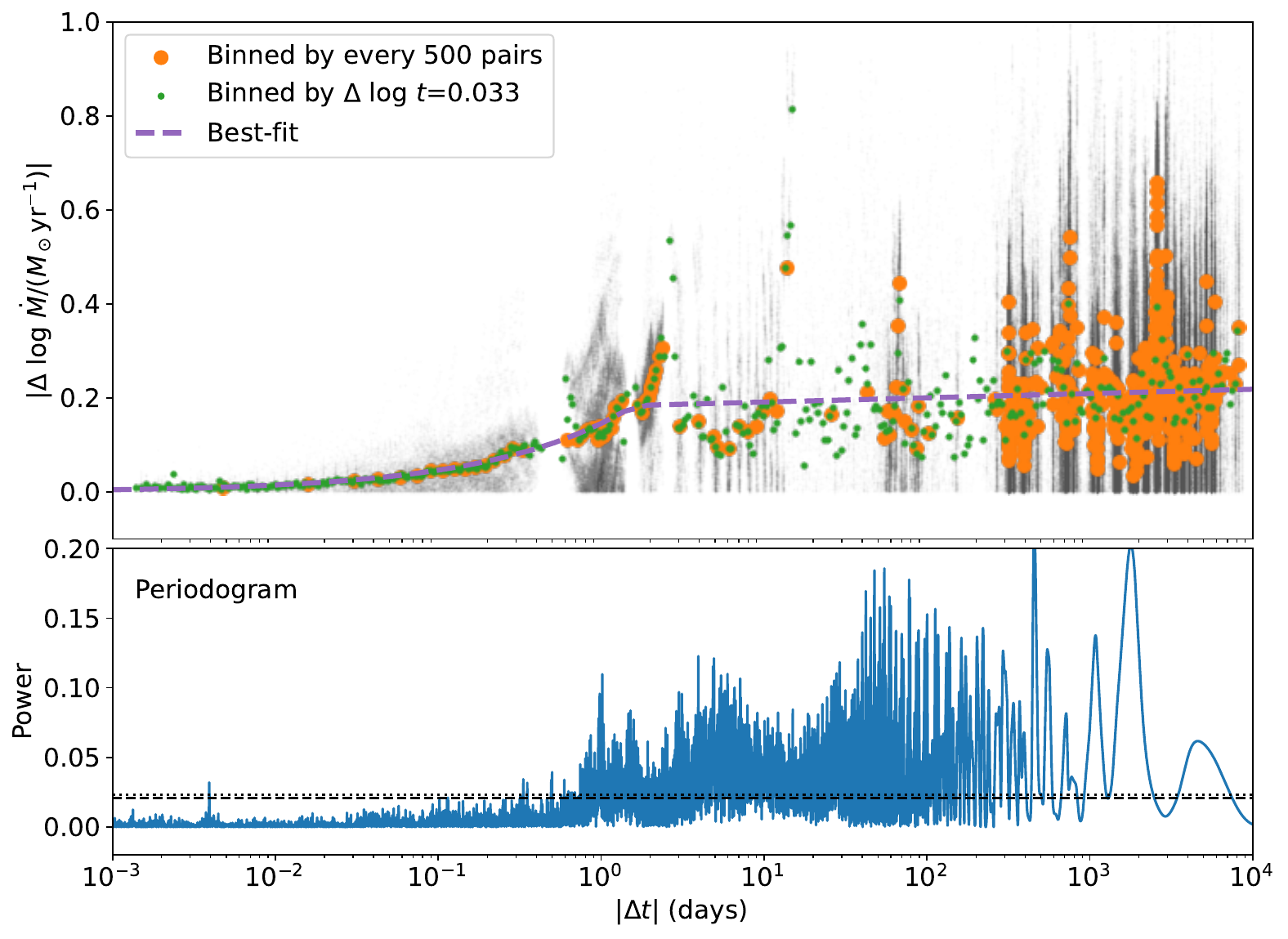}
\caption{ \textit{Top}: The so-called ``$\Delta m$-$\Delta t$'' diagram of the $\log \dot M$ variabilty. Each black point on the background represents a pair of $\dot M$ measurements, with x values being the absolute difference in time ($|\Delta t|$), and y values being $|\Delta \log \dot M|$. Individual pairs are bins in the order of $|\Delta t|$ by every 500 pairs (orange points) or by every $|\Delta t| = 0.033$ (green points). The purple dashed line is the two-part function that fits the binned points best, with a break at $\sim 1.6$ days from a power-law increase to a flat line (see Equation \ref{eq:two-part} and Table \ref{tab:twopart} for details).  \textit{Bottom}: The periodogram of the $\log \dot M$ variability. The horizontal dashed and dotted lines represent the 5\% and 1\% false-alarm rate respectively. }
\label{mdotvary}
\end{figure*}

\subsection{The reset timescale for accretion rate}

To evaluate any characteristic timescales of accretion, we first focus on a robust determination of the nonperiodic variations using structure functions.  The difference in accretion rates is compared with the time separation for every two datapoints in our sample. This so-called ``$\Delta m$-$\Delta t$'' analysis is  widely used to study time variations, including for quasars \citep[e.g][]{deVries2005}, for a similar spectroscopic analysis of a young accreting star, DF Tau, by \citep{johnskrull97}, and also on larger samples of accretion rates and photometry by \citet{Costigan2014}, \citet{zsidi22}, and \citet{sergison20}.  A detailed description of structure functions and their limitations for young stars is provided by 
\citet{Findeisen2015}. A monotonic increase of $\Delta m$, the difference in physical property, with respect to $\Delta t$, the time difference, such that the system is further away from the start point as time passes, indicates that the system is in a varying mode. However, if the $\Delta m$-$\Delta t$ relation is flat, it suggests that the system relaxes and the variation occurs on a shorter timescale.  

For our dataset, the distribution of differences in accretion rates, $\Delta \log \dot M$, is given as,
\begin{equation}
|\Delta \log\dot M|=|\log \dot M_1-\log\dot M_2|=|\log\frac{\dot M_1}{\dot M_2}|,
\end{equation}
where $\dot M_1$ and $\dot M_2$ are the accretion rates of each pair of data points as a function of the time difference between those two data points. Despite the large diversity of timescales in this dataset, we do not have enough high-cadence data  at short timescales to reliably distinguish the difference between increasing $\dot M$ and decreasing $\dot M$, hence the absolute value of $|\Delta \log \dot M|$ is used. 

Figure \ref{mdotvary} shows the $|\Delta \log \dot M|$-$|\Delta t|$ distribution of the 692,076 pairs from the 1169 $\dot M$ measurements. The median $|\Delta \log \dot M|$-$|\Delta t|$ in bins of $|\Delta t|$ are presented in orange (binned by every 500 pairs) and green (binned by $\Delta \log t = 0.033$) points. At short timescales ($\lesssim 1$ day), the median $|\Delta \log \dot M|$ increases consistently with $|\Delta t|$. However, on long timescales ($\gtrsim 1$ day), the median value flattens with respect to $|\Delta t|$ and is nearly flat at $\sim 0.2$ dex, despite a large spread in the $|\Delta \log \dot M|$ values. This analysis leads to the same level of variation as the distribution of accretion rates presented in Figure~\ref{fig:histogram}, with the additional conclusion that the accretion rates are (usually) independent from one day to the next.
To quantify the turnaround timescale, we fit the median binned $|\Delta \log \dot M|$-$|\Delta t|$ relations, from two binning methods combined, with a two-part function, 
\begin{equation}
\label{eq:two-part}
|\Delta \log \dot M| = 
\begin{cases}
  A_0 |\Delta t|^\gamma,  & |\Delta t| < \tau_0 ;\\    
  A_1 \log \left( \frac{|\Delta t|}{\tau_0} \right) + A_0 \tau_0^\gamma, & |\Delta t| \ge \tau_0, 
\end{cases}
\end{equation}
where $A_0, A_1, \gamma$, and $\tau_\mathrm{0}$ are set as free parameters. For the equation, the unit of $\dot M$ is $M_\odot~\mathrm{yr}^{-1}$ and the units for $\Delta t$ and $\tau_0$ are days. The best-fit results are listed in Table \ref{tab:twopart}.

\begin{table}[!b]
\centering
\caption{Parameters of the two-part function}
\begin{tabular}{lc}
\hline
Name$^a$ &  Best-fit \\
\hline
$A_0$ & $0.15 \pm 0.01$\\
$A_1$ & $0.009 \pm 0.003$ \\
$\gamma$ & $0.50 \pm 0.08$ \\
$\tau_0$ & $1.6 \pm 0.4$ \\
\hline
\multicolumn{2}{l}{$^a$ See Equation \ref{eq:two-part} for definitions.}\\
\end{tabular}
\label{tab:twopart}
\end{table}

According to the fitting result, most of the accretion variability occurs on a timescale within $\tau_0 = 1.6 \pm 0.4$ days. On longer timescales, the accretion rate appears randomly distributed around the mean value.
However, the $|\Delta \log \dot M|$-$|\Delta t|$ relation is not entirely flat, with a positive slope ($A_1 = 0.009 \pm 0.003$) that suggests the mean accretion rate slowly (either incoherently or coherently) drifts away from the starting point. 

\citet{Findeisen2015} found that the time scale in the $\Delta m$-$\Delta t$ plot is reliable when the time value is $\sim 30$ times greater than the sampling interval and $\sim 1/15$ of the total time baseline.  The sampling of our data has a minimum interval = 0.008 days and a time baseline of $\sim 25$ years. Therefore, the 1.6-day time scale is generally robust.  %The time sampling at periods shorter than one day is small but sufficient to be statistically robust.

TW Hya has been previously found to have periodic (or quasi-periodic) behavior with 2--4.5 day periods, attributed to a combination of accretion hot spots and dark starspots on the stellar surface \citep[e.g.][]{Mekkaden1998,Lawson2005,Huelamo2008,Rucinski2008}.  Possible periodic signals could occur on short timescales related to oscillations of the accretion flow \citep{sacco10}, rotational timescales, turbulent timescales \citep{robinson21}, or a viscous timescale on the order of $\sim 2$ months \citep{takasao22}.

We search for these periods in accretion rate by using the generalized Lomb-Scargle peridograms \citep{lomb76,scargle82,zechmeister09}.  The lower panel of Figure~\ref{mdotvary} shows the periodogram with the false-alarm probability (FAP) estimated from bootstrap. At high-frequencies (period $T \lesssim 1$ day), most power is within the background level with low significance, except a peak with FAP $ < 1\%$ at $T \sim 0.004$ days. We examine this signal with phase-folding diagrams and find no convincing periods. Furthermore, the period, which corresponds to $\sim 5$~min, is roughly the cadence of individual exposures for consecutive observations. Therefore, we conclude that this signal is likely due to the sampling window function of the observations.  At lower frequencies ($T \gtrsim 1$~day), the power spectrum becomes noisy with a few prominent peaks close to the observational cadences of 1 day, 29 days, and 365 days. Therefore, despite the power spectrum is consistently above the 1\% FAP, the signal is likely due to the combination of these observational cadences.

\subsection{Interpreting the timescales for variable accretion onto TW Hya}

High-frequency photometric monitoring of TW Hya \citep[e.g.][]{siwak18} reveals a lightcurve littered with constant accretion bursts and decays, so many of these timescales may relate to the rise and decay of these bursts.  Frequent stochastic bursts are expected for accreting young stars, based on simulations of the magnetospheric geometry \citep{romanova08,Blinova2015,takasao19}.  We see only individual points within those stochastic lightcurves, though in a few cases the veiling lightcurve covers many consecutive hours \citep[e.g.][]{Dupree2012}.

Although accretion fluctuates on any given day, over the long term the average accretion rate appears stable.  The structure function indicates that accretion resets on $\sim 1.6$ day timescales.  Similar analyses from photometry of larger samples indicate that $\tau_0=0.25 \times P_*$, where $P_*$ is the rotation period \citep{sergison20,venuti21}, which for TW Hya would have suggested a reset timescale of $\sim 0.9$ days.

In analysis of high time-resolution photometry, \citet{Siwak2011} found power on a timescale of 1.3 days, which may be a consequence of the timescale for the reset of accretion rates on the 1.6 day timescale found here.
 This 1.6 day timescale approximates how quickly the accretion rate changes with time, but this change is not periodic. 
 The stellar dipole field is roughly constant, at least on short timescales \citep[e.g.][]{Donati2011}, however, the amount of gas lifted off the disk and accreted onto the star may be constantly changing because of asymmetric structures in the inner disk.  The $1.6$ day timescale for resetting accretion is similar to the timescales obtained from structure functions in other protoplanetary disks \citep[e.g.][]{venuti21,zsidi22}.

\section{Conclusions}
\label{sec_conclusion}

We measure the veiling in 1169 high-resolution spectra of TW Hya obtained over 25 years.  These measurements are converted to accretion rates based on scalings obtained from 26 flux-calibrated low-resolution spectra.  
From analyzing this dataset, we find the following results:

\begin{enumerate}

\item The veiling at 5000-5100 \AA\ varies from 0.13 to 2.13, equivalent to mass accretion rates from $0.47-9.43 \times10^{-9}$ M$_\odot$ yr$^{-1}$, with an average of $2.51\times10^{-9}$ M$_\odot$ yr$^{-1}$.  These accretion rates are only slightly lower than the average accretion rate for accreting stars in nearby young clusters.  With its current mass, the disk could survive for another 3.8 Myr.  The accretion rate would be 12\% higher if we adopted the truncation radius of 3.5 R$_*$ measured by \citet{gravity_garcialopez20}.

\item The distribution of accretion rates is well described by $\log \dot{M}_{\rm acc}=-8.65$ with a standard deviation of 0.22 dex.  Accretion never ceases. This variability is less than the scatter seen in accretion rate-stellar mass relationships \citep{manara22}.  The variability is consistent with the level of accretion variability of solar-mass stars in many other sources \citep[e.g.][]{venuti21,zsidi22}, but with some important exceptions, such as XX Cha \citep{claes22}.

\item The uncertainty in our veiling measurements is $<0.05$ ($\sim0.003$ when comparing broadband spectra from the same instrument), which is a minor contribution to the overall error budget.  The uncertainty in accretion rates is instead dominated by bolometric corrections and to a lesser extent spots. Comparisons to measurements from \citet{robinson19} indicate that the inclusion of multiple accretion streams at lower densities could increase the accretion rate by as much as $\sim 50\%$.  We also find that the accretion spectrum becomes bluer when the accretion rate is higher, which leads to a bolometric correction that depends on accretion rate.  This dependence is not incorporated into our accretion rate measurements.

\item He lines correlate reasonably well with the accretion luminosity.  However, the H$\alpha$ luminosity and 10\% width are only weakly correlated with accretion luminosity because of higher opacities and absorption in accretion streams and winds and because of possible temporal offsets between continuum and line emission.  The use of H$\alpha$ and H$\beta$ as accretion rate indicators for TW Hya is not recommended.  These correlations also fail for XX Cha \citep{claes22} and should be used with caution, although at least some sources show strong correlations between accretion and line emission \citep[GM Aur,][]{bouvier23}.
 Long-term variability is found in the [\ion{O}{1}] $\lambda6300$ line, which traces the inner disk and MHD wind.

\item The line emission is excluded from the calculated accretion rates and are an important correction for TW Hya.  The exclusion of lines, and particular Ly$\alpha$ emission, from the accretion luminosity would increase the average accretion rate by 70\% to $4.3\times10^{-9}$ M$_\odot$ yr$^{-1}$.  
The exclusion of lines such as Ly$\alpha$ emission may lead to systematic underestimates of all accretion rates.  Including the lines in instantaneous accretion rate estimates would decrease the scatter in accretion rates.

\item The $\Delta m$-$\Delta t$ diagram indicates that the reset timescale for the aperiodic accretion variability is $\sim 1.6$ days.  Large increases in accretion are likely caused by short bursts, as seen in photometry \citep{siwak18}. We did not find convincing evidence of periodic signal associated with the bursts. Despite the 1.6-day timescale for resetting accretion rate, some epochs appear to have different average accretion rates.  Accretion onto TW Hya was weaker than average in Feb.~2014 and stronger than average in Mar-Apr. 2022. 

\end{enumerate} 

The accretion variability of TW Hya is likely common and unremarkable in the context of other accretion disks, but this lengthy time series provides new insights into unremarkable variability.  The steady accretion likely occurs along dipole-like field lines that land near the pole \citep[e.g.][]{Donati2011}, with high-velocity emission and absorption in He lines that indicate mass loading onto these field lines during bursts.  While the accretion rate appears unstable on any given day, the overall accretion rate and fluctuations in that rate have been stable over the past 25 years.

\section{Acknowledgements}

We thank the referee for a careful read and comments, which saved us from publishing a significant error and misinterpretation of possible periods in the veiling.
We thank the many people who have contributed time and effort into obtaining this data at the telescope, reducing the data, and developing archives to make the data available.  This list includes Ilaria Pascucci, Feng Long, and Claudio Melo.  We also thank Kevin France for discussions of Ly$\alpha$ emission.  This work benefited from discussions with the ODYSSEUS team (HST AR-16129; \citealt{espaillat22}, \url{https://sites.bu.edu/odysseus/}).

GJH and JT are supported by the National Key R\&D program
of China 2022YFA1603102 and by general grant 12173003 from the National Natural Science Foundation of China.
ZG acknowledges support from FONDECYT Postdoctoral 3220029.
ZG acknowledges support by ANID Millennium Science Initiative Program NCN19\_171.
Support for this HMG was provided by the National Aeronautics and Space Administration through Chandra Award Number GO1-22007X issued by the Chandra X-ray Observatory Center, which is operated by the Smithsonian Astrophysical Observatory for and on behalf of the National Aeronautics Space Administration under contract NAS8-03060.  AF and JA acknowledge support by the PRIN-INAF 2019 STRADE (Spectroscopically TRAcing the Disk dispersal Evolution) and by the Large Grant INAF YODA (YSOs Outflow, Disks and Accretion).
This work has been funded by the European Union under the European Union Horizon Europe Research \& Innovation Programme 101039452 (WANDA). This project has received funding from the European Research Council (ERC) under the European Union's Horizon 2020 research and innovation programme under grant agreement No 716155 (SACCRED).  JFD acknowledges funding from the European Research Council (ERC) under the H2020 research and innovation programme (grant 740651, NewWorlds). SHPA acknowledges financial support from CNPq, CAPES
and Fapemig.  JHK is supported by NASA XRP grant 80NSSC19K0292 and NASA ADAP grant 80NSSC22K0625 to RIT.

This work is based in part on observations obtained at the Canada-France-Hawaii Telescope (CFHT) which is operated by the National Research Council of Canada, the Institut National des Sciences de l´Univers of the Centre National de la Recherche Scientique of France, and the University of Hawaii.  The CFHT observations were obtained in programs 08AF11, 10AP11, 12Ap12, 14AP18, and 16AP18 (PI Donati) and 15BE97 (PI Malo).

This work is also based in part on observations collected at the European Southern Observatory under ESO programmes 106.20Z8 (PI Manara), 074.A-9021 (PI Setiawan), 089-A-9007, 090.A.9013, 092.A-9007 (PI Mohler), 093.A-9029 (PI Gredel), 099.A-9010 (PI Sarkis), 099.A-9008 (PI Mueller), 0101.A-9012 (PI Launhardt), 075.C-0202 (PI Gunther), 081.C-0778, 082.C-0390 (PI Weise), 082.C-0427 (PI Doellinger), 082.C-0218 (PI Melo), 089.C-0299 (PI Pascucci), 085.C-0238 (PI Alcala), 085.C-0764 (PI G\"uenther), 103.200T (PI G\"unther), and 60.A-9036 and 60.A-9022 (engineering runs, no PI listed). 

This research is based on observations made with the NASA/ESA Hubble Space Telescope obtained from the Space Telescope Science Institute, which is operated by the Association of Universities for Research in Astronomy, Inc., under NASA contract NAS 5-26555.  These observations are associated with programs GO-8041 (PI Linsky), GO-9093 (PI Johns-Krull), 11608 (PI Calvet), and 13775 (PI Espaillat) and can be accessed via \dataset[http://dx.doi.org/10.17909/0v8a-vq31].

Some of the data presented herein were obtained at the W. M. Keck Observatory, which is operated as a scientific partnership among the California Institute of Technology, the University of California and the National Aeronautics and Space Administration. The Observatory was made possible by the generous financial support of the W. M. Keck Foundation.  This research has made use of the Keck Observatory Archive (KOA), which is operated by the W. M. Keck Observatory and the NASA Exoplanet Science Institute (NExScI), under contract with the National Aeronautics and Space Administration.  The Keck data were obtained in programs C199LA (Herczeg), C247Hr (PI Carpenter), C252Hr (PI Hillenbrand), C186Hr (PI Hillenbrand), C199Hb (PI Herczeg), adnd C269Hr (PI Dahm).

Observing time with SMARTS/Chiron obtained by PI Walter was made possible by a Research Support grant from Stony Brook University.
We thank Wei-Chun Jao, Leonardo Paredes, and Todd Henry for managing the Chiron spectrograph and their prompt scheduling of the requested observations.

The authors wish to recognize and acknowledge the very significant cultural role and reverence that the summit of Maunakea has always had within the indigenous Hawaiian community.  We are most fortunate to have the opportunity to conduct observations from this mountain.

Views and opinions expressed are however those of the author(s) only and do not necessarily reflect those of the European Union or the European Research Council. Neither the European Union nor the granting authority can be held responsible for them.

\section{Appendix A: High and low veiling accretion spectra}

\begin{table}[!t]
\centering
\caption{Coadded spectra$^a$}
\begin{tabular}{lccc}
\hline
Spectrum & \# & $r_{5050}$ range & avg$(r_{5050})$ \\
\hline
$f0$ &  14 & 1.1--1.52 & 1.25 \\
$f1$ &  17 & 0.80--0.85 & 0.83\\
$f2$ &  12 & 0.60--0.65 & 0.62\\
$f3$ &  11 & 0.4--0.45 & 0.43 \\
$f4$ &  11 & 0.2--0.25 & 0.23 \\
$f5$ &  17 &$-0.01$--0.03 & 0.01\\
\hline
\multicolumn{4}{l}{Relative veiling values listed.}\\
\end{tabular}
\label{tab:coaddspectra}
\end{table}

In data behind the figures, we provide high S/N spectra by coadding 10--15 spectra of similar veiling, including the high and low veiling spectra and the residual spectrum in Figure~\ref{fig:lines_hilo}. 
These high and low veiling spectra were identified after initial veiling measurements, and then veiling was remeasured to produce the residual spectrum.
For each of these spectra, the number of coadds, veiling range, and veiling average are provided in Table~\ref{tab:coaddspectra}. The spectra are coadded over every pixel by normalizing the flux over $\sim 30$ \AA\ around each pixel. 

The residual spectrum is calculated by first measuring the veiling from 5000--5100, $r_{5050}$ in the high accretion spectrum, $f0$, and the low accretion spectrum, $f5$.  We then apply the correlations between the veiling at other wavelengths and $r_{5050}$ (as shown in Figure~\ref{fig:veilcomp}), to obtain a veiling spectrum.  The veiling spectrum is then converted to flux (following methods in Figure~\ref{fig:veilspec}), so that excluded wavelengths could be estimated by extrapolating from nearby wavelengths.  We then reconvert this flux spectrum into a veiling spectrum, tailored to the photospheric flux level.  Finally, we subtract the low veiling spectrum from the high veiling spectrum, after scaling the low veiling spectrum to its estimated contribution to the high veiling spectrum.

The residual spectrum is what remains after this process.  Photospheric lines are no longer present in the spectrum.  Instead, the spectrum is mostly the flat accretion continuum.  However, many narrow emission lines are also present across this spectrum.  For many of these lines, the emission fills in the core of the photospheric lines and would not be detectable in a single spectrum, but the excess emission is apparent after subtracting off a high quality photospheric template, in this case TW Hya itself during weak accretion epochs.

\begin{table}[!t]
\centering
\caption{Equivalent widths of photospheric lines used for veiling}
\begin{tabular}{cccccc}
\hline
$\lambda_{obs}$ (\AA) & EW (\AA) &  $\lambda_{obs}$ (\AA)& EW (\AA) & $\lambda_{obs}$ (\AA)& EW (\AA) \\
\hline
     4591.587 &      0.116 &    5426.490 &      0.128 &    5737.296 &      0.093 \\
   4592.824 &      0.122 &    5434.755 &      0.229 &    5857.699 &      0.197 \\
   4594.290 &      0.138 &    5455.876 &      0.249 &    5866.697 &      0.116 \\
   4611.438 &      0.122 &    5490.386 &      0.082 &    5942.003 &      0.100 \\
   4613.517 &      0.139 &    5497.736 &      0.242 &    5956.953 &      0.090 \\
   4617.468 &      0.106 &    5501.700 &      0.130 &    6013.738 &      0.097 \\
   4651.480 &      0.128 &    5506.979 &      0.180 &    6016.900 &      0.113 \\
   4881.779 &      0.117 &    5535.707 &      0.100 &    6020.359 &      0.081 \\
   4903.494 &      0.086 &    5569.851 &      0.092 &    6022.045 &      0.114 \\
   4934.267 &      0.142 &    5573.104 &      0.165 &    6065.740 &      0.119 \\
   4965.066 &      0.116 &    5582.192 &      0.146 &    6081.698 &      0.102 \\
   4991.293 &      0.192 &    5588.984 &      0.174 &    6085.493 &      0.108 \\
   5007.420 &      0.096 &    5590.333 &      0.153 &    6090.467 &      0.103 \\
   5189.038 &      0.109 &    5594.700 &      0.220 &    6111.902 &      0.103 \\
   5219.918 &      0.107 &    5601.519 &      0.137 &    6141.977 &      0.126 \\
   5252.306 &      0.094 &    5603.094 &      0.184 &    6150.414 &      0.099 \\
   5261.925 &      0.116 &    5671.083 &      0.096 &    6154.478 &      0.092 \\
   5262.455 &      0.110 &    5672.044 &      0.086 &    6166.693 &      0.102 \\
   5282.017 &      0.098 &    5682.867 &      0.154 &    6199.438 &      0.108 \\
   5341.218 &      0.173 &    5688.446 &      0.156 &    6210.933 &      0.112 \\
   5346.033 &      0.247 &    5698.711 &      0.165 &    6216.628 &      0.114 \\
   5348.536 &      0.187 &    5703.818 &      0.089 &    6219.547 &      0.087 \\
   5349.724 &      0.190 &    5707.230 &      0.094 &    6231.034 &      0.148 \\
   5353.646 &      0.080 &    5709.677 &      0.118 &    6246.586 &      0.081 \\
   5394.902 &      0.153 &    5727.291 &      0.121 &    6252.094 &      0.111 \\
   5415.437 &      0.114 &    5727.900 &      0.091 &    6261.393 &      0.126 \\
   5420.589 &      0.124 &    5731.489 &      0.108 &    6274.934 &      0.093 \\
   6285.420 &      0.092 \\
   \hline
\end{tabular}
\label{tab:lineeqws}
\end{table}

\section{Appendix B: Relationships between veiling diagnostics and final veiling measurements}

This section describes how equivalenth widths and spectral indices are used to calculate veiling, with measurements from ESpaDonS spectra.  Table~\ref{tab:lineeqws} lists 82 photospheric absorption lines that are coadded for equivalent width measurements.  Each line is normalized by the nearby continuum before being added to other nearby lines.  Table~\ref{tab:eqwtoveil} describes the lines that are coadded and the relationship between equivalent width and veiling.  The lines that are used and coadded differ, depending on spectral coverage and resolution.  

Veiling is also measured using the spectral indices in Table~\ref{tab:indices}.  Each spectral index is calculated by dividing the median flux in a low-flux region by the median flux of a nearby region with stronger emission.  The relationship between each spectral index and veiling is then described with a 4th-order polynomial fit to ESPaDOnS data.

\begin{table}[!th]
\centering
\caption{Relationships to convert equivalent widths to veiling}
\begin{tabular}{lcccc}
\hline
&  \multicolumn{4}{c}{$\sum a_i \times$ EW$_i$}\\
Line range (\AA) & $a_0$ & $a_1$ & $a_2$ & $a_3$ \\
\hline
4500--5200 &     -0.758  &   -5.920 &    28.355  &   62.400\\
5200--5450 &    -1.783 &   -21.250  &  -68.016  & -144.046\\
5450--5600 &    -1.451  &  -20.997  &  -81.493  & -280.907\\
5600--5800 &     -1.922 &   -19.083 &   -54.677 &  -132.374\\
5800--6100 &     -2.540  &  -26.538 &   -95.976 &  -198.139\\
6100--6300 &     -3.970  &  -48.499 &  -208.793 &  -379.800 \\
\hline
\end{tabular}
\label{tab:eqwtoveil}
\end{table}

\begin{table*}[!th]
\centering
\caption{Spectral indices and veiling}
\begin{tabular}{lcccccc}
\hline
Region of & Region of & \multicolumn{4}{c}{$\sum a_i \times (F_{low}/F_{cont})^i$}\\
Low Flux (\AA) & Continuum Flux (\AA)& $a_0$ & $a_1$ & $a_2$ & $a_3$ & $a_4$\\
\hline
 & 5186--5188.5, 5190--5191.5\\
5166--5185$^a$ & 5175--5158.5, 5160.0-5162.5,5164--5165 & 331.36 & -1848.60 &  3855.37 & -3568.97  & 1240.63\\
\hline
& 5196.7--5200, 5212--5219.5\\
5204--5211.5 & 5220.5--5222, 5240--5242 &     -20.04  &  106.60  & -208.50   & 170.49  &  -43.19\\
\hline
 & 5285.5--5294, 5256.5--5260\\
5262--5276.5 & 5277.5--5279.5,5245--5246.5 &    1153.54 & -5924.35 & 11409.70 & -9775.75  & 3147.71\\
\hline
5323-5330.5$^b$ & 5312--5316,5335--5338 &    -200.01  &  837.71 & -1245.03  &  739.73  & -126.41\\
\hline
5369--5373 & 5354--5361,5374.5--5375.5,5378--5382 &     301.27 & -1700.42  & 3572.08 & -3324.61 &  1160.96\\
\hline
5396.5--5411.3$^c$ & 5388-5392.5,5401--5403.5,5416.5--542 &     543.57  &-2928.97  & 5911.19 & -5305.77  & 1790.89 \\
\hline
6160.5--6171 & 6144--6146,6171.5--6173,6178--6179.5 &     577.36  &-3480.02  & 7849.18  &-7864.79  & 2959.92\\
\hline
\multicolumn{6}{l}{$^a$excluding 5169--5170.5~~~~~$^b$excluding 5325--5327~~~~~$^c$excluding 5398--5404, 5407--5409}\\
%\multicolumn{3}{l}{excluding 6164.5--6169}\\
\end{tabular}
\label{tab:indices}
\end{table*}

\clearpage
\pagebreak

\bibliographystyle{apj}
\bibliography{ms}

\end{CJK*}
\end{document}